\documentclass[nofootinbib,aps,amssymb,preprint,tightenlines,letterpaper,showkeys]{revtex4}

\usepackage{graphicx,psfrag}

\renewcommand{\a}{\alpha}

\newcommand{\G}{\Gamma}
\renewcommand{\d}{\delta}\newcommand{\D}{\Delta}
\newcommand{\e}{\epsilon}
\newcommand{\f}{\phi}\newcommand{\F}{\Phi}

\renewcommand{\i}{\imath}

\renewcommand{\v}{\nu}

\renewcommand{\O}{\Omega}
\newcommand{\p}{\psi}
\renewcommand{\r}{\rho}
\newcommand{\s}{\sigma}\renewcommand{\S}{\Sigma}
\renewcommand{\t}{\theta}
\newcommand{\X}{\chi}

\newcommand{\cC}{{\mathcal C}}
\newcommand{\cD}{{\mathcal D}}
\newcommand{\cE}{{\mathcal E}}

\newcommand{\cH}{{\mathcal H}}

\newcommand{\cL}{{\mathcal L}}
\newcommand{\cM}{{ M}}

\newcommand{\cO}{{\mathcal O}}

\newcommand{\cS}{{\mathcal S}}
\newcommand{\cT}{{ T}}

\newcommand{\cV}{{\mathcal V}}

\newcommand{\br}[1]{\bar{#1}}
\newcommand{\td}[1]{\tilde{#1}}

\newcommand{\da}{\daleth}
\newcommand{\bra}{\langle}
\newcommand{\ket}{\rangle}

\newcommand{\te}{\td{e}}

\newcommand{\tG}{\td{G}}
\newcommand{\tD}{\td{\D}}
\newcommand{\wD}{\tilde{\D}}

\newcommand{\ctT}{\td{{\cal T}}}

\newcommand{\be}{\begin{equation}}
\newcommand{\ee}{\end{equation}}
\newcommand{\bea}{\begin{eqnarray}}
\newcommand{\eea}{\end{eqnarray}}
\newcommand{\bD}{\br{\D}}

\renewcommand{\le}{\left}
\newcommand{\ri}{\right}

\newcommand{\R}{\mathbb{R}}
\newcommand{\C}{\mathbb{C}}
\newcommand{\N}{\mathbb{N}}
\newcommand{\Z}{\mathbb{Z}}

\newcommand{\SU}{\mathrm{SU}}

\renewcommand{\det}{\mathrm{det}}
\newcommand{\tr}{\mathrm{tr}}

\newcommand{\prodor}[1]{\overrightarrow{\prod_{{#1}}}}

\newcommand{\stackindex}[2]{\begin{array}{c}\vspace{-8mm} {}_{{#2}}
\\ {}_{{#1}} \end{array}}

\newcommand{\mm}[4]{\left(\begin{array}{cc}
{#1} & {#2} \\ {#3} & {#4}
\end{array}\right)}

\newcommand{\ox}{\otimes}

\newcommand{\wmod}[1]{\stackrel{({#1})}{{\cal V}}}
\newcommand{\wrep}[1]{\stackrel{({#1})}{\Pi}}
\newcommand{\zjt}{Z_{\D,T,T^*}(\{j_e\},\{\t_{e^*}\})}
\newcommand{\ljt}{L_{\D,\cT,\cT^*}(\{j_e\},\{\t_{e^*}\})}

\newcommand{\eval}{\mathrm{eval}}

\newtheorem{prop}{Proposition}
\newtheorem{theo}{Theorem}
\newtheorem{lemma}{Lemma}
\newtheorem{defin}{Definition}
\newcommand{\proof}{\textit{Proof : }\\}

\newcommand{\figeq}[1]{\begin{array}{c} {#1}\end{array}}

\begin{document}

\begin{titlepage}
\title{\large Ponzano-Regge model revisited II:
Equivalence with Chern-Simons}
\author{ Laurent Freidel}
\email{lfreidel@perimeterinstitute.ca}
\affiliation{\vspace{2mm}Perimeter Institute for Theoretical
Physics\\ 35 King street North, Waterloo  N2J-2G9,Ontario,
Canada\\} \affiliation{Laboratoire de Physique, \'Ecole Normale
Sup{\'e}rieure de Lyon \\ 46 all{\'e}e d'Italie, 69364 Lyon Cedex
07, France }

\author{David Louapre}
\email{dlouapre@ens-lyon.fr} \affiliation{\vspace{2mm}Laboratoire
de Physique, \'Ecole Normale Sup{\'e}rieure de Lyon \\ 46
all{\'e}e d'Italie, 69364 Lyon Cedex 07, France}\thanks{UMR 5672
du CNRS}

\date{\today}

\begin{abstract}
We provide a mathematical definition of the gauge fixed
Ponzano-Regge model showing that it gives a measure on the space
of flat connections whose volume is well defined. We then show that
the Ponzano-Regge model can be equivalently expressed as
Reshetikhin-Turaev evaluation of a colored chain mail link based
on $\cD(\SU(2))$: a  non compact quantum group being the Drinfeld
double of $SU(2)$ and a deformation of the Poincare algebra. This
proves the equivalence between spin foam quantization and
Chern-Simons quantization of three dimensional gravity without
cosmological constant. We extend this correspondence to the
computation of expectation value of physical observables and
insertion of particles.
\end{abstract}
\maketitle
\end{titlepage}

\tableofcontents

\section{Introduction}
The Ponzano-Regge model was the first model of quantum gravity ever
proposed \cite{PR}. It gives a prescription for the computation of
scalar product in 3dimensional Euclidean quantum gravity.
 It is a state sum model over discretized
geometries where the main building block is the Racah-Wigner 6j
symbol of the classical $SU(2)$ group and it has been mainly
studied by physicists. The asymptotic of the 6j symbol was
noticed to be given by the cosine of the Regge action \cite{PR}.
Arguments in favor of this asymptotic were given \cite{Nev} and a
mathematical proof appeared in \cite{Ro} (see also \cite{FL6j} for
a geometrical simple proof). The relation between the
Ponzano-Regge model and 3d gravity was further understood: It was
proven that the Ponzano-Regge model realizes a projection on the
space of flat $SU(2)$ connections \cite{Ooguri} and that it can be
obtained from a discretization of a first order formulation of
gravity (\cite{FK} and reference therein).

On the mathematical side, there is the Turaev-Viro model
\cite{TV}, which is the analog of Ponzano-Regge model but based on
the quantum group $SU(2)_q$  with q a root of unity $q=\exp
(i2\pi/k)$. For this model, the sum over representations is finite
and the invariance under refinement of the triangulation is exact
and not formal. Moreover, it is well known that classically Euclidean 2+1
gravity with a positive cosmological constant is equivalent to a
$SO(4)$ Chern-Simons theory \cite{Witt}. This
equivalence has been proven to extend at the quantum level: Not
only the Turaev-Viro partition function is the square of the
$SU(2)$ Chern-Simons partition function \cite{Turo,sktv},
 but there is also a one to one correspondence between
the states of Turaev-Viro's Hilbert space and Chern-Simons gauge
invariant operators \cite{Kar,ConfFK}. This correspondence makes
it clear that the Turaev-Viro model is related to Euclidean 3d gravity with
a cosmological constant $\Lambda = (2\pi/k)^2$. This is consistent
with
the asymptotic of $SU(2)$ 6j symbols \cite{Mizo,Wood}. However,
one should regret that there is not yet, a direct relationship
between this model and a discretization of gravity as this is the
case for the Ponzano-Regge model. Also, the relation between the
Turaev-Viro and Ponzano-Regge model is not well understood: One
would naively expect that the limit when $k$ goes to $\infty$ of
the Turaev-Viro model to reproduce the Ponzano-Regge amplitude.
This is not true since the representations of spin comparable to
the cut-off $k$ still contributes in the limit \cite{Wood2,dual}.
Eventually, 3d Euclidean or Lorentzian Gravity with or without
cosmological constant is always equivalent classically to a Chern-Simons theory.
However, it is only in the case of Euclidean gravity with positive cosmological constant,
the case address by the Turaev-Viro model,
that the Chern-Simons gauge group is compact $SO(4)\sim SU(2)\times SU(2)$.
 In the other cases, Lorentzian gravity
or Euclidean gravity with $0$ or negative cosmological constant the gauge group
is either a Poincare group or
$SO(3,1)\sim SL(2,\C)$ or $SO(2,2)\sim SL(2,\R)\times SL(2,\R)$ hence a non-compact gauge group.
It is then necessary, if we want to understand quantum Lorentzian geometry,
to have a well defined mathematical model describing $3d$ quantum gravity
based on a noncompact gauge group.
The Ponzano-Regge model is the simplest model we know addressing
this question\footnote{see \cite{Lorlau} for a state sum model
describing Lorentzian 3d gravity}.

Therefore, we still need to have a direct mathematical
understanding of the Ponzano-Regge model. This is the purpose of
this paper. Two problems need to be addressed:
 First, the Ponzano-Regge was ill defined as a mathematical
model. There are divergences in the summation that need to be
regularized, moreover, after regularization, the Ponzano-Regge
model is not necessarily invariant under refinement of the
triangulation \cite{Bar}. It was only recently understood
\cite{FLdiffeo, FLDSU2phys} that these residual divergences were due to a
residual `translational' symmetry of the model. This symmetry
being non compact should  be gauge fixed. In this paper we review
the definition of the properly gauge fixed Ponzano-Regge model
from a mathematical point of view. We show that, given a three
manifold $M$, the Ponzano-Regge model defines a measure on the
space of flat $SU(2)$ connections on $M$ and that the volume of
this measure is truly independent of the triangulation and gauge
fixing.

The second aspect of the theory addressed here is the equivalence
with Chern-Simons theory at the quantum level. At the classical level, 2+1 gravity with
no cosmological constant is equivalent to Chern-Simons theory
based on the Poincare group. It was shown in \cite{FLDSU2phys}
that there is an hidden quantum group symmetry in the
Ponzano-Regge model: The braiding of particles being represented
by the action of the braiding matrix of $\cD(\SU(2))$. This is a non
compact quantum group which is the quantum double of $SU(2)$ and a
kappa deformation of the Poincare group \cite{bais2, FLeeJerzy}. This quantum
group is also known to be the building block of the combinatorial
Chern-Simons quantization of 2+1 gravity without cosmological
constant \cite{Schroers}. In this article, we show that the
Ponzano-Regge partition function can be written as the
Reshetikhin-Turaev evaluation, based on $\cD(\SU(2))$, of a chain mail
link. This proves the equivalence at the quantum level between the
spin foam and Chern-Simons quantization.
 We also
prove that the evaluation of Chern-Simons Wilson lines reproduces
the insertion of spinning particles which was proposed in
\cite{FLDSU2phys}.

We have included in the main text our definitions and
main results. In order to simplify the reading we have put proofs of
our statements in appendix for the more dedicated reader. We have
also  included a review of old and new  results on the recoupling
theory of $\cD(\SU(2))$.

The plan of the paper is as follows: In section \ref{reginv} we
give the definition of the regularized (gauge fixed) Ponzano-Regge
invariant, and justification for this definition. We show that
this invariant is the volume of  the space of flat $SU(2)$
connection modulo gauge transformation with respect to measure
depending on the triangulation and gauge fixing. We also exhibit a
relationship between the singularity structure of this space and
potential divergence of this volume. We present one of our main
theorem stating the independence of this volume under change of
triangulation and gauge fixing.

In section \ref{sec:CM} we present the construction of the chain
mail link associated to a 3d manifold and the chain mail link
invariant associated with $\cD(\SU(2))$. We then prove the equivalence
between this link invariant and the Ponzano-Regge model. The
quantum group being non compact, one has to carefully define such
link invariant using special properties of the braiding matrices.
We also present key properties such as sliding identities of the chain mail
evaluation and give a chain mail proof of the independence under gauge fixing.

In section \ref{sec:obs} we show the equivalence between
expectation value of  Ponzano-Regge observables, like the Wilson lines or particle observables
\cite{FLDSU2phys},
and $\cD(\SU(2))$
evaluation of links. We use this correspondence to give an
alternative chain mail proof of the independence of our invariant
under change of gauge fixing.

In appendix \ref{app:su2}, \ref{app:dsu2} we recall well known facts about
$SU(2)$ and $\cD(\SU(2))$.
The appendix \ref{app:braidings} contains all the key technical new results concerning the
evaluation of $\cD(\SU(2))$ links and $\cD(\SU(2))$ fusion identities.
In order  to simplify the reading we have put in
appendix \ref{lemma1}, \ref{app:proofinv}  the proofs of the main lemma and theorem of section
 \ref{reginv}

\section{The gauge fixed Ponzano-Regge invariant}\label{reginv}
 In this section, we define an invariant
that actually corresponds to a gauge fixed version of the
Ponzano-Regge model \cite{FLdiffeo,FLDSU2phys}.
 We will not present here the physical
analysis that motivates this model but present instead a
independent mathematical justification of our definition.

\subsection{Definitions and notations}\label{def}
 The invariant
is defined using a triangulation $\D$ of a connected three
dimensional compact manifold $\cM$ with boundary $\partial
M$\footnote{If the boundary is not empty we suppose for technical
reason that none of its connected component is a 2-sphere, so
$\sum_i g_{\Sigma_i}>0$ where $g_{\Sigma_i}$ is the genus of
$\Sigma_i$}. $\D$ induces a 2d triangulation of the boundary
denoted $\bD \equiv \partial \D$. We will also denote by $\D_i$
(resp. $\bD_i$) the i-skeleton of $\D$ (resp.  $\bD$) consisting
of open cells of dimension $i$. We will refer to the internal
vertices, edges and faces (triangles) of this triangulation by
$v\in \D_0,e\in \D_1,f\in \D_2$ and the boundary ones by
$\bar{v},\bar{e},\bar{f} \in \bar{\D}$.

\begin{figure}[ht]
\begin{minipage}[t]{0.96\linewidth}
\begin{minipage}[c]{0.46\linewidth}
\includegraphics[width=0.70\linewidth]{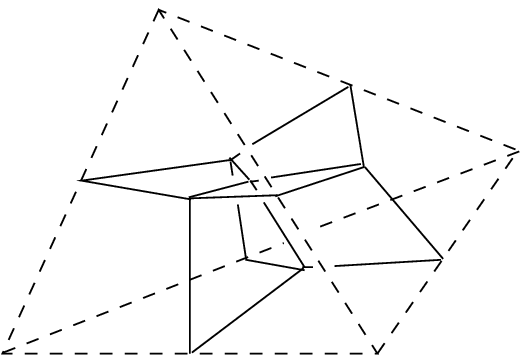}
\end{minipage}
\begin{minipage}[c]{0.46\linewidth}
\includegraphics[width=0.50\linewidth]{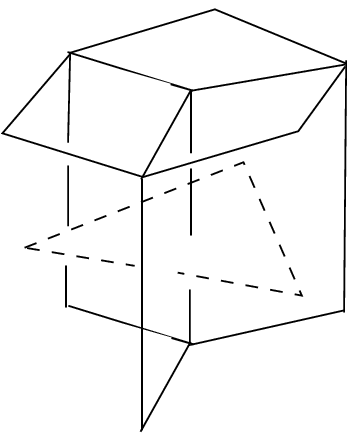}
\end{minipage}
\end{minipage}
\caption{Dual complex around a tetrahedra and a boundary
triangle.}\label{Dual}
\end{figure}

We also need to introduce the cellular complex $\wD^{*}$ dual to
the triangulation $\D$. This complex consists of two components:
an interior component $\D^{*}$ and a boundary component $\bD^{*}$.
The interior component is denoted $\D^{*}$ and its i-skeleton
$\D^{*}_{i}$ is in one to one correspondence with $ \D_{3-i}$. The
union of the 0 and 1 skeleton of $\D^{*}$ is a  graph with open
ends. Its internal vertices are barycenters of tetrahedra, its
internal edges are dual to faces sharing two tetrahedra and its
open ends are dual to boundary triangles. The boundary part of
$\tD^{*}$ is denoted $\bD^{*}$ and is dual in $\partial \cM$ to
the boundary triangulation. There is an isomorphism $\bD^*_i
=\bD_{2-i}$. The union of the 0 and 1 skeleton of $\bD^{*}$ is a
trivalent graph whose vertices are dual to boundary triangles. The
full dual complex is the union of $\bD^{*}$ with $\D^{*}$, where
the open ends of $\D^{*}$ are glued to the vertices of $\bD^{*}$.
It is denoted by $\wD^{*}$ and there is an isomorphism $\wD^*_i
=\D_{3-i} \cup \bD_{2-i}$. We will refer to the internal vertices,
edges and faces of the dual triangulation by $v^*\in
\D_0^{*}¥,e^*\in \D_1^{*}¥,f^*\in \D_2^{*}¥$ and the boundary ones
by $\bar{v}^*,\bar{e}^*,\bar{f}^* \in \bar{\D}$ This construction
is illustrated in the figure \ref{Dual}.

 In order to formulate the model, we will
also need additional ingredients. First, we chose an orientation,
denoted $or$, of the edges and faces of $\D$. This orientation
induces an orientation of the dual faces and dual edges of $\D^*$.
We denote by $\e(f^*,e^*)=\pm 1$ the relative orientation of $f^*$
and $e^*$ so $ \partial f^* = \sum_{e^{*}\subset f^{*}}¥
\e(f^*,e^*) e^*$.
 Also, for each dual face $f^*\in \widetilde{\Delta}^*_2$,
we pick up a dual vertex
in it, that we call starting dual vertex of $f^*$ and we denote it
$st(f^*)$.

We  also need to a chose an internal maximal tree $\cT \subset \D$
and a dual
 maximal tree $\cT^*\subset \wD^*$.
A tree of a graph is a subgraph containing no loop.
A tree is said to be maximal if it is touching all the vertices of the
graph and is connected.
We choose $\cT^{*}$ a maximal tree  of $\widetilde{\D}^{*}_{1}$.
We  choose $\cT$  to be a connected tree
of $\D_{1}$ which touches every
internal vertex of  $\D_{1}$ and touches  $\partial \cM$
in only one vertex of $\bD_{1}$.

We denote this additional ingredients by $d\equiv(or,st,T,T^*)$
and denote $\D_d$ the triangulation decorated by these data.
Given $\D_d$  we can first equipped the triangulated
manifold with a discrete connection.
A discrete connection is an assignment of
group elements $g_{e^{*}}\in G=\SU(2)$ to any oriented dual edge
$e^{*}\in \wD^{*}$.
A change of dual edge orientation amounts to take the inverse group
element: $g_{-e^{*}}=g_{e^{*}}^{-1}$.
A discrete gauge group $G^{|\D_{0}^{*}|}$ is acting
on this connection at the vertices of the dual triangulation
\be
g_{e^{*}} \rightarrow  k_{t_{e^{*}¥}}g_{e^{*}}k_{s_{e^{*}¥}}^{-1},
\ee
$s_{e^{*}},t_{e^{*}}$ being the source and target of the edge $e^{*}$.

Given a maximal tree $\cT^{*}$ and using the gauge group action, we
can fix the value of $g_{e^{*}}$ to be the identity for all edges
${e^{*}} \in \cT^{*}$. This gauge fixed connection $g^{\cT^{*}}$
is an element of $G^{\wD_{\cT^{*}}}$ where $ \wD_{\cT^{*}}
=\wD^{*}_1 \backslash \cT^{*}$ is the set of edges of $ \wD^{*}$
not belonging to $T^*$. The residual gauge group action on a gauge
fixed connection is the diagonal adjoint action $ g_{e^{*}} \to  k
g_{e^{*}}k^{-1}$.

Given a discrete connection we can assign
to any  face $f^{*}\in \wD_{2}^{*}$,
an holonomy
\begin{equation}\label{eqn:defcurvature}
G_{f^*}=\overrightarrow{\prod_{e^*\subset f^*}} g_{e^*}^{\e(e,e^*)}.
\end{equation}
The product is understood to be taken along the edges $e^*$ in
the dual face $f^*\sim e$, starting from $st(f^*)$ and according
to the orientation of $f$. $\e(f^*,e^*)=\pm 1$ denotes the relative
orientation of $e$
and the one induced by the orientation of $f\sim e^*$.
If $f^{*}$ is in $ \D_{2}^{*} \subset  \wD_{2}^{*}$ it is dual to an
edge $e \in \D_1$ and in this case we will denote $G_e$ the holonomy
around
$f^* \sim e$.

The space of flat connections is denoted $\cal{A}_{F}$ and  defined
to be  the sets of connection $g_{e^{*}}$  for which
$G_{f^{*}} =1$ for {\it all} dual faces $f^{*}$.
A flat connection gives a representation of  $\Pi_{1}(\cM)$ in $G$
so ${\cal{A}_{F}}= Rep(\Pi_{1}(\cM),G)$.

Given a decorated triangulation $\D_d$ with $d= (or,st,T,T^*)$ we
have a map
\bea\label{map}
G_{\D_{d}} : G^{\wD_{\cT^{*}}} &\rightarrow & G^{\D_{\cT}} \\
(g_{e^{*}})_{e^{*}\in \wD_{\cT^{*}} } &\rightarrow & (G_{e})_{e \in
\D_{\cT} }
\eea
where $\D_{\cT} = \D_1 \backslash \cT$ is the set of
edges of $\D$ not in $\cT$.
We then have the lemma
\begin{lemma}
   \be {\cal{A}_{F}}= G_{\D_{d}}^{-1}(1),\ee
   and
   \be\label{diffg}
   |\wD_{\cT^{*}}|-|\D_{\cT}|= 1+ \sum_{\Sigma_{i}\in \partial \cM¥}
   (g_{\Sigma_{i}}-1),
   \ee
   where the sum is over all  connected components of $\partial
\cM$\footnote{if
   there is no such component the RHS of (\ref{diffg}) is $0$}
   and $g_{\Sigma_i}>0$ denotes the  genus of each connected component
   $\Sigma_i \subset \partial M$.
   If there is no such component the RHS of (\ref{diffg}) should be
$0$
 \end{lemma}
 This lemma states that the conditions $G_{\D_{d}} =1$
 is a sufficient set of conditions implying $G_{f^{*}} =1$ for
 all $f^*$, even the ones not in $\D_T$.
 The second part of the lemma suggest that the conditions
 $G_{\D_{d}} =1$ are generically not redundant.
 The LHS of (\ref{diffg}) computes the number of generators
 of $\Pi_1(M)$
 minus the number of relations we impose on this generators.
 The RHS of (\ref{diffg}) gives an  bound on the
 minimal dimension of $ \Pi_1(M)$:
 $dim(\Pi_1(M))\geq 1+ \sum_i (g_{\Sigma_{i}}-1) $
 or $dim(\Pi_1(M))\geq 0$ if $\partial M=\emptyset$.
 This bound is saturated for some manifolds. For instance
 when the manifold is closed and
 $\Pi_{1}(\cM)$ is a discrete set
or when $M$ is a genus $g$ handlebody $H_{g}$
 in which cases it is known that
 $dim(\Pi_{1}(H_{g}))= g $.
 The proof of this lemma is presented in appendix \ref{lemma1}.

The previous  lemma motivates the following definition.
\begin{defin}[Gauge fixed Ponzano-Regge invariant]
   Given a triangulation $\D_d$ of a manifold $\cM$,
equipped with an orientation $\e$ of the edges and faces of $\wD$,
and a choice $st(f^*)$ of starting dual vertex for dual
face and a choice of a maximal tree $\cT$ and a dual maximal tree
$\cT^*$,
we assign groups elements $g_e^*\in G$ to oriented dual edges of
$\wD^{*}$. We
consider the quantity
\begin{equation}\label{eqn:defregulPR}
PR[\cM_{\D_d}] = \int \prod_{e^*\in \wD^{*}} dg_{e^*}
\le(\prod_{e^*\in\cT^*} \d(g_{e^*}) \ri)\ \le(\prod_{e\in\Delta
\backslash\cT}
\d(G_e)\ri).
\end{equation}
The measure $dg$ is the normalized
Haar measure of $G=\SU(2)$ and $\delta(g)$ denotes the corresponding
Dirac distribution
on the group.
\end{defin}

In the following, we prove that this definition is actually an
invariant of the 3-manifold $\cM$ not depending essentially on the
triangulation additional data.
When it is finite this invariant computes the volume of the moduli
space of flat $SU(2)$
connection on $M$.

But first, lets focus on the definition which contains integrals over
products of distributional delta function. The integrals can be
potentially divergent
and in this case $PR$ is infinite.

In order to understand this issue lets consider the following
distribution
\be\label{deD}
\delta_{\D_{d}}(x_{e})
\equiv \int \prod_{e^*\in \wD^{*}} dg_{e^*}
\le(\prod_{e^*\in\cT^*} \d(g_{e^*}) \ri)\ \le(\prod_{e\not\in\cT\cup
\bD}
\d(G_e x_{e}^{-1})\ri).
\ee
The Ponzano-Regge invariant is defined as the value of this
distribution
at $x_e=1$.
In general the value of a distribution at a given point is ill-defined
unless the distribution is sufficiently regular at the identity.
Lets suppose that the data $d=({\D,\cT,\cT^*,or})$ is such that  the
volume
distribution $\delta_{\D_{d}}$ is regular  (continuous or of class
$C^k$).
Under this condition it can be written as a limit\footnote{If $f$ is
a bounded
continuous function one can write $f(x)=\lim_{\e\to 0}\int
\d_{\e}(xg^{-1})f(g)dg$}
\be\label{de}
\delta_{\D_{d}}(x_{e})=\lim_{\epsilon \to
0}\delta_{\D_{d}}^{(\e)}(x_{e})=
\lim_{\epsilon \to 0} \int \prod_{e^*\in \wD^{*}\backslash \cT^*}
dg_{e^*}
\le(\prod_{e\in\Delta \backslash \cT }
\d_\e(G_e x_{e}^{-1})\ri),
\ee
where $\d_\e(g)$ is the heat kernel on $\SU(2)$ solution of
$\partial_\e \d_\e(g)= \D_{G}¥ \d_\e(g)$, with $\D_{G}$ the
invariant
Laplacian on the group. Since this is a $C^\infty$ class function
it can be expanded it terms of character as
$\d_\e(g)=\sum_{j}d_j e^{-\e c_j} \chi_j(g)$, where $d_j$ is the
dimension,
$\chi_j(g)$ the characters and $c_j$ the quadratic Casimir of the
spin $j$ representation.
The integral (\ref{de}) can be compactly written as
\be
\delta_{\D_{d}}^{(\e)}(x_{e})=\int_{G^{\wD_{\cT^{*}}}}
\delta_\e(G_{\D_{d}}(g)x^{-1}) dg,
\ee
where $G_{\D_{d}}$ is the map (\ref{map}).
When $\e \to 0$ the integral localizes on  $G^{-1}_{\D_{d}}(x)$.
More precisely, lets denote by
$S\subset G^{\wD_{\cT^{*}}}$ the set of critical points
of $G_{\D_{d}}$\footnote{$x \in X$ is said to be a critical point of
$G: X \to Y$,
where $\mathrm{dim}(X) >\mathrm{dim}(Y)$,
if $ d_xG$ is not of maximal rank.
$y \in Y$ is said to be a singular value if $G^{-1}(y)$ contains
critical
points,
otherwise it is a regular value.
By Sard theorem the set of critical value is
of measure $0 $}.
If $S$ is of measure $0$ in $G^{\wD_{\cT^{*}}}$, then
$\int_{G^{\wD_{\cT^{*}}}} \delta_\e(G_{\D_{d}}(x)) dx=
\int_{G^{\wD_{\cT^{*}}}\backslash S} \delta_\e(G_{\D_{d}}(x)) dx$
and taking the limit $\e \to 0$ localizes the integral \cite{liu, sengupta}
\be\label{locaint}
\delta_{\D_{d}}(x_{e})=\int_{G_{\D_{d}}^{-1}(x_{e})\backslash S}
\frac{dx}{|\det(d_xG_{\D_{d}})^*|},
\ee
where $d_xG_{\D_{d}}$ denotes the differential of $G_{\D_{d}}$ at $x$
and $(d_xG_{\D_{d}})^*$ the dual linear map.
Since the set $S$ is by definition the locus of the zero
of $|\det(d_xG_{\D_{d}})^*|$, this integral is indefinite
and potentially infinite when $x_{e}$ is not a regular value.
When $x_e=1$ the integral localizes
over the moduli space of flat $SU(2)$ connection ${\cal A}_F=
G_{\D_{d}}^{-1}(1)$.
This space is not a manifold when the identity is a singular value of
$G_{\D_{d}}$.
However the space ${\cal A}_F^0 ={\cal A}_F \backslash S$ is a
(non-compact) smooth manifold. The Ponzano-Regge invariant computes,
when $\delta_{\D_{d}}$ is regular at the identity,
the volume of this manifold with the induced measure (\ref{locaint}).

Lets suppose $M$ is an  homology sphere (a closed manifold). In
this case the trivial connection is isolated and let suppose that
${\cal A}_F$ contains irreducible connections. By definition, the
gauge group acts non trivially on irreducible connections.
Therefore they are critical points of $G_{\D_{d}}$. In this case
the Ponzano-Regge invariant is infinite and in general we do not
expect this invariant to be finite for topologically non
trivial closed manifold.
 In the case of open manifold
we know, on the contrary, some examples where the Ponzano-Regge
invariant is well defined \cite{FLDSU2phys}. It is a very
interesting problem to understand under which topological
conditions the gauge fixed PR invariant is finite. It is also tempting
to ask wether one can extend our definition to give a well defined
value of the Ponzano-Regge invariant for  all manifolds. For
instance we can use the heat kernel regularization (\ref{de}) to
extract from $\delta_{\D_{d}}^{(\e)}(1)$ a finite part. The
challenge is then to prove that such a procedure leads to a result
which is independent of the triangulation as we show in the next
section for our definition of. These questions are  beyond the
scope of our paper and  we will not be addressed here.

\subsection{The invariance theorem}

To prove that this definition actually leads to an invariant of
the manifold $\cM$, we have to prove that it is independent of the
choice of all additional ingredients we introduced.

\begin{theo}\label{theo:PRinv}
The definition of $PR[\cM_{\D_d}]$ is  independent of:
The choice of the orientation of the $e$, of the orientation
of the $f$, of the choice of the $st(e)$.
The choice of the maximal trees $\cT$ and $\cT^*$: It depends only on
the
homology class of $T,T^*$\footnote{ This is  called a combinatorial
Euler structure.
In \cite{Tur} Turaev relates combinatorial Euler structures to smooth
Euler
structures
given by a nowhere vanishing vector field on $M$ normal to the
boundary}.
The choice of the triangulation $\D$ of $M$.
It is thus an invariant of $\cM$ that we will denote $PR(\cM)$.
\end{theo}
\proof{The detailed proof of this theorem is given in appendix
\ref{app:proofinv}.
 The independence of the choice of orientation and starting
vertices is based on the properties of the Haar measure and
$\d$-function that insure invariance under reversing orientations
or changing start vertices. This proof is given into proposition
\ref{prop:invorient}.

The proof of the independence of choice of trees relies on the
fact that two homologous trees can always be related by a sequence of
elementary moves (described appendix \ref{app:proofinv} lemma
\ref{lemma:movetrees}). We then prove that the properties of the
integral definition (\ref{eqn:defregulPR}) are such that two
expressions obtained for trees related by an elementary move are
actually the same. This is presented in propositions
\ref{prop:invtree} and \ref{prop:invtreestar}.

The proof of the independence of the choice of $\D$ are done by
proving the invariance under Pachner moves. A Pachner move lead
from a triangulation $\D$ to a triangulation $\tD$. To compute
(\ref{eqn:defregulPR}) defined with $\tD$, we need to use a
maximal tree of $\tD$. The proof goes as follows : we choose
maximal trees $\cT,\cT^*$ for $\D$ and compute
$PR(\cM,\D,\cT,\cT^*)$. From these trees we explicitly construct
maximal trees $\tilde{\cT},\tilde{\cT}^*$ for $\tD$ and compute
$PR(\cM,\tD,\ctT,\ctT^*)$ to prove that we obtain the same result.
This is the result of proposition \ref{prop:invtriang}.
}

\subsection{State-sum expression of the Ponzano-Regge invariant}

In this section we introduce the notion of amplitude associated to a
\textit{colored}
triangulation. A colored triangulation is obtained by assigning a
half-integer to each edge of $\D$ and element of $H/W=[0,2\pi]$ to
each face (dual edge), where $H/W$ denotes the Cartan subgroup of
$\SU(2)$ quotiented by its Weyl group.

\begin{defin}\label{def:zjt}
Let us consider a triangulation $\D$ with trees $T,T^*$, colored by
$j_e$ for each edge
$e\not\in T$
and $\t_{e^*}$ for dual edge $e^*\notin T^*$. We
associated to this colored triangulation an amplitude
\begin{equation}\label{eqn:zjtdef}
\zjt=\le(\prod_{e^*\not\in\cT^*} \int_{G/H} dx_{e^*} \ri)
\prod_{e\not\in\cT} \X_{j_e}(\prodor{e^*\subset e}
x_{e^*}h^{\e(e^*,e)}_{\t_{e^*}}x_{e^*}^{-1}),
\end{equation}
where $\X_{j_e}$ is the character of the spin $j_e$
representation, and $h_{\t_{e^*}}\in SU(2)$,  is the representative
element
of $\t_{e^*} \in U(1)$ given in appendix \ref{app:su2}.
\end{defin}
We have seen in the previous section that
The Ponzano-Regge invariant is given
by the evaluation of the  functional
$\delta_{\D_d}(x_e)$ at the identity.
Using repeatedly the Plancherel formula
\begin{equation}
\d(G)\sim \sum_j d_j \X_j(G),
\end{equation}
with $d_j =2j+1$ is the dimension of the spin $j$ representation.
We can compute the Fourier expansion of this functional and get
\be\label{fourier}
\delta_{\D_d}(x_e) \sim \sum_{j_e}d_{j_{e}}
\le(\prod_{e^*\not\in\cT^*} \int dg_{e^*} \ri)
\prod_{e\not\in\cT} \X_{j_e}(\prodor{e^*\subset e}
g_{e^*}^{\e(e^*,e)}x_e).
\end{equation}
This is an equality in the sense of distributions but this is not
in general a pointwise equality between functions. If we restrict
to a so called {\it regular} choice of $M$ and
 $\D$ for which $\delta_{\D_d}(x_e)$ is regular (continuous) at the
identity we have $\delta_{\D_d}(x_e)= \lim_{\e \to 0}
\delta_{\D_d}^{(\e)}(x_e)$. Since $\delta_{\D_d}(x_e)$ is a smooth
function it is equal to its Fourier series. We can use the Weyl
integration formula (cf appendix \ref{app:su2}) to decompose the
group integrations
\begin{equation}
\int_G dg f(g)=\int_{0}^{2\pi} \frac{d\t}{\pi} \D^2(\t) \int_{G/H} dx
f(xh_{\t}x^{-1}),
\end{equation}
where $\D(\theta)= \sin(\frac{\theta}{2})$ and
where $dg$ and $dx$ are normalized invariant measures on $G$ and
$G/H$.
We then have the following proposition :
\begin{prop}[Ponzano-Regge invariant as a state
sum]\label{prop:PRstatesum} The Ponzano-Regge invariant of a
regular manifold $\cM$ is given by
\begin{equation} \label{eqn:PRsumdef}
PR(\cM_d)=\lim_{\e \to 0} \le(\prod_{e\not\in\cT} \sum_{j_e}e^{\e -c_j}
d_{j_e}\ri)\
\le(\prod_{e^*\not\in\cT^*} \int_0^{2\pi} \frac{d\t_{e^*}}{\pi}
\D^2(\t_{e^*})
\ri)\ \ \zjt.
\end{equation}
\end{prop}
If the sum over $j_e$ at $\e=0$ is absolutely convergent we can
exchange
the summation and limit process and forget about the $\e \to 0 $,
limit.
In the following we will restrict to the study of this case.

Note that if we perform the integrations over $g_{e^*}$
in (\ref{fourier}) or the $\t_{e^*}$ integration in
(\ref{eqn:PRsumdef}) and uses recoupling identity
one obtains the invariant in the  Ponzano-Regge form
\be
PR(M) = \lim_{\e \to 0} \sum_{j_e}e^{-\e c_j}d_{j_{e}}\prod_{e\in T}
\delta_{j_e,0}
\prod_{t\in \D_3}
\prod_{t}
\left\{
\begin{array}{ccc}
    j_{e_{t_{1}}} &  j_{e_{t_{2}}} &  j_{e_{t_{3}}} \\
    j_{e_{t_{4}}} &  j_{e_{t_{5}}} &  j_{e_{t_{6}}}
    \end{array}
    \right\},
    \ee
where the summation is over all edges of $\D$ and the product is
over all tetrahedra, the weight associated to each tetrahedra is
the normalized 6j symbol and $e_{t_{i}}$ denotes the six edges
belonging to the tetrahedra $t$.
$d_j=2j+1$ denotes the dimension of the spin $j$ representation and
$\delta_{j_e,0}$
denotes the Kronecker delta symbol.


\section{Chain-mail invariant using $\cD(\SU(2))$}\label{sec:CM}

The Turaev-Viro invariant has been proved by Roberts to be equal
to its chain-mail invariant, built using Reshetikhin-Turaev
evaluation for the quantum group $U_q(\SU(2))$. In this part we
propose to built the analogue of this equivalence for the
Ponzano-Regge model, but using the evaluation with quantum double
$\cD(\SU(2))$.

\subsection{Construction of the chain-mail link $L_\D$}
\subsubsection{Closed manifold}
\begin{figure}[ht]
\begin{minipage}[t]{0.96\linewidth}
\begin{minipage}[c]{0.46\linewidth}
\includegraphics[width=0.35\linewidth]{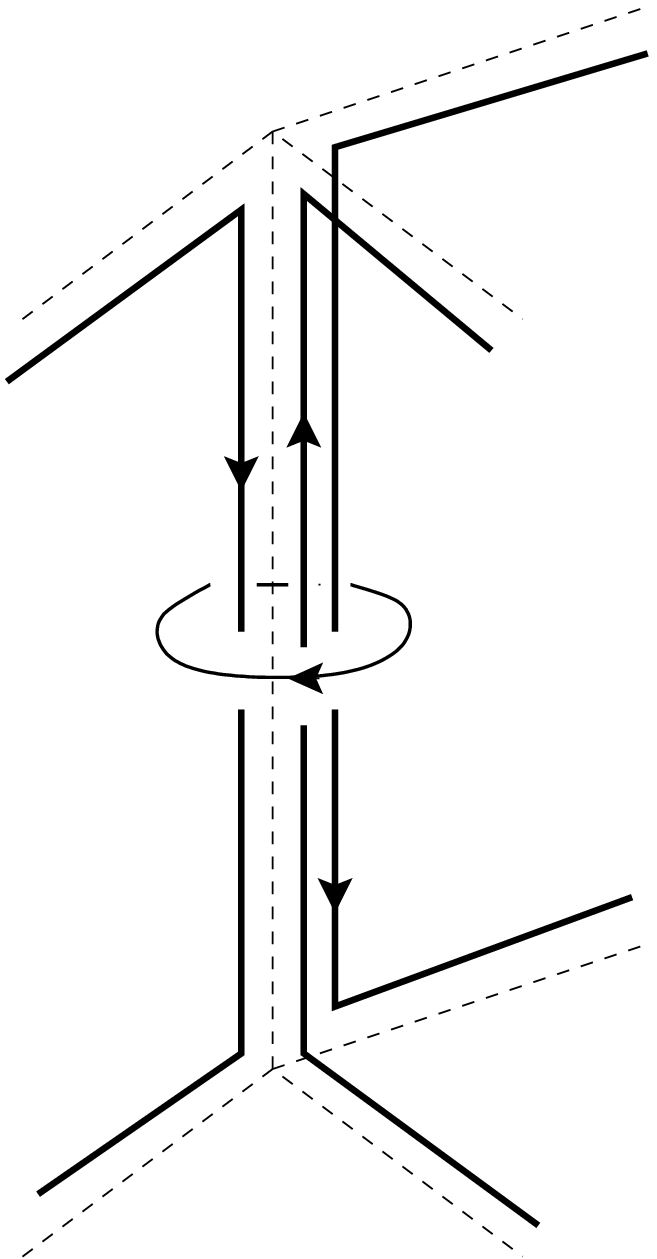}
\end{minipage}
\begin{minipage}[c]{0.46\linewidth}
\includegraphics[width=0.50\linewidth]{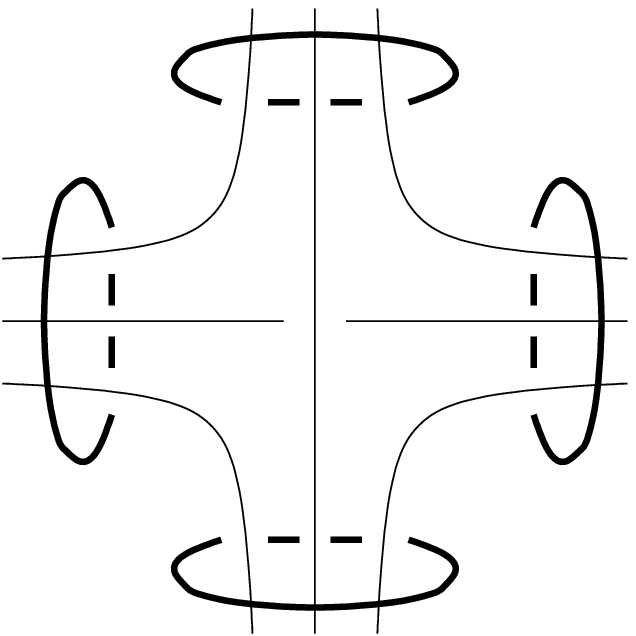}
\end{minipage}
\end{minipage}
\caption{Construction of chain-mail link around an edge and
resulting picture for a tetrahedra.}\label{fig:chainmail}
\end{figure}

We explain here the construction by Roberts \cite{sktv} of the
chain-mail link $L_\D$ associated to a triangulation $\D$ of a
closed manifold $\cM$. Consider a closed manifold $\cM$ and a
triangulation $\D$ of it, equipped with orientation of edges and
faces and a choice of a starting vertex for all dual faces
$f^*\sim e$. To each face  $f^*\sim e$ of the dual  complex
${\D}^*$, we associate a circle component $C_e$ which is the
boundary of $f^*$ slightly push inside $f^*$. To each edge of the
dual complex $\D^*$ we associate a circle $C_{e^*}$ which is the
boundary of a small disk transverse to  the edge $e^*$ and
encircling all the $C_e$ components associated with dual faces
touching $e^*$ (see figure \ref{fig:chainmail}).
 We also equipped each component of the link with
an orientation provided by the one of the corresponding face or
edge. Finally, the starting vertex $st(f^*)$ is encoded in the
marking of a point on each component $C_{e^*}$. The union of $C_e$
and $C_{e^*}$ is the chain mail link $L_{\D}$, see figure
\ref{fig:chainmail}. Given trees $T,T^*$ we construct the Chain
mail link $L_{\D,T,T^*}$ by deleting all the $C_e$ and $C_{e^*}$
components for $e\in T$, $e^*\in T^*$.

A more geometrical way to describe the construction of the link
$L_\D$ uses a Heegard splitting of $\cM$. Such a splitting can
be obtained from the triangulation $\D$ in the following way : we
consider the 1-skeleton of $\D$ and thicken it to obtain a
handlebody $H$ of boundary $\S$. Now the dual 1-skeleton of $\D$
can be also thicken to get another handlebody $H^*$ whose boundary
is also $\S$. The boundaries of $H$ and $H^*$ are naturally
identified and this gives a Heegard splitting
\begin{equation}
\cM=H \#_{\S} H^*.
\end{equation}
The link $L_\D$ is built from this splitting by considering the
meridians of $H$ and $H^*$. These meridians are in one to one
correspondence with edges of $\D_{1}$ and $\D^{*}_{1}$.
We draw this meridians and equipped
them with a framing given by thickening them inside the boundary
$\S$, and with an orientation induced by the corresponding
orientation of the edges and faces of the triangulation. Pushing
slightly the meridians of $H^*$ into $H$ gives a linking between
all of these meridians. One thus get an oriented framed link
$L_\D$ which has the following properties.
\begin{enumerate}
\item Each of its components is an unknot.
\item It is made of two different types of components, the
$C$-components, being the meridians of $H$ and the $C^*$
components being the meridians of $H^*$.
\item By construction, components of the same type are not linked
together.
\end{enumerate}

Given  the chain-mail link $L_\D$ and a
choice of maximal trees $\cT$ and $\cT^*$ in the 1-skeleton and
dual 1-skeleton of $\D$, the link
$L_{\D,\cT,\cT^*}$  is obtained by removing
from $L_\D$ the components coming from meridians around edges of
$\cT$ and dual edges of $\cT^*$.
If we denote by $V, E, F, T$
respectively the number of vertices, edges, faces and tetrahedra
in $\D$. A maximal tree of the triangulation is by construction
made of $V$ vertices and $V-1$ edges, while a dual maximal tree is
made of $T$ vertices and $T-1$ edges. The number of $C$-components
of $L_{\D,\cT,\cT^*}$ is thus $E-V+1$ while the number of
$C^*$-components is $F-T+1$. For a closed orientable 3-manifold,
these two numbers are equal since its Euler characteristic is zero
\begin{equation}
\X(\cM)=V-E+F-T=0.
\end{equation}
In this case the link determines an element $F$ of the mapping
class group of $\partial H$ which is such that under $F$ the
sublink of $L_\D$ made of the $C$ components is mapped to the
sublink made of the $C^*$ component. $F$ is uniquely determined up
to multiplication on the left and right by automorphism of $H$.

\subsubsection{Open manifold}
\begin{figure}
\psfrag{Ce}{$C_e$} \psfrag{C*}{$C^*_e$}
\includegraphics{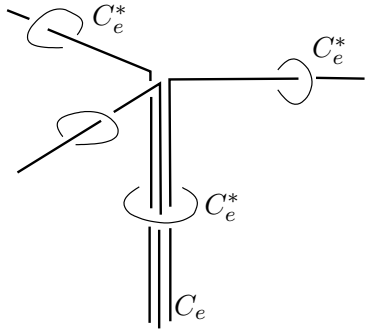}
\caption{Construction of the chain-mail link near the
boundary.} \label{fig:boundCM}
\end{figure}

The chain mail link can be also defined in the case of a manifold
with boundary as follows:
To each face  $f^*\sim e$ of the dual interior complex ${\D}^*$,
we associate a circle component $C_e$ which is the boundary of $f^*$
slightly
pushed inside $f^*$.
To each edge of the full dual complex $\tilde{\D}^*$
we associate a circle $C_{e^*}$ which is the boundary of a
small disk transverse to  the edge $e^*$ and encircling all the
$C_e$ components.
The union of $C_e$ and $C_{e^*}$ is the Chain mail link
$L_{\D}$, see figure \ref{fig:boundCM}.

Given trees $T,T^*$ we construct the Chain mail link
$L_{\D,T,T^*}$ by annihilating all the $C_e$ and $C_{e^*}$ components
for
$e\in T$, $e^*\in T^*$.
All the $C_{e}$ components are marked by the choice of a staring
vertex
$st(f^*)$.

\subsection{$\cD(\SU(2))$ and its Reshetikhin-Turaev evaluation}

The general idea of the Reshetikhin-Turaev \cite{Turo} evaluation of a colored
link using quasi-triangular Hopf algebra is the following. We
consider a quantum group acting  on
representation spaces $V_\r$. An oriented framed colored link
$L(\{\rho_C\})$ is given by the assignment of representations
$\rho_C$ of the quantum group to every component $C$ of an
oriented framed link $L$. If one chose a point on each component
of the link we can cut the link along this points and obtain a
tangle which contains no closed loops. The original link is
recovered as the closure of the tangle.
 The Reshetikhin-Turaev evaluation
associates to every colored tangle an endomorphism of
$\bigotimes_b V_{\rho_{C_b}}$ where $b$ denote the strands of the
tangle and $C_b$ the corresponding color. This endomorphism is
obtained using the $R$-matrix structure of the quantum group and the
ribbon element and gives an invariant of framed tangle.
 The evaluation of the colored link
$L(\{\rho_C\})$ is then obtained from here by taking the (quantum)
trace of this endomorphism and leads to a link invariant.

 Our goal is
to use the Reshetikhin-Turaev evaluation with the quantum group
$\cD(\SU(2))$. Basic facts about the quantum group $\cD(\SU(2))$ are
summarized in appendix \ref{app:dsu2}. This quantum group is
constructed from $SU(2)$ by using the Drinfeld double construction
and is therefore a Ribbon Hopf algebra. We recall here few facts
about its representation theory. The irreducible unitary
representations of $\cD(\SU(2))$ are labelled by a couple $(\t,s)$
where $\t\in[0,2\pi[$ and $2s\in\N$. We will denote the
corresponding representations by $\wrep{\t,s}$ and the carrier
spaces of representation by $\wmod{\t,s}$. These carrier spaces
are infinite dimensional if $\theta \neq 0$ and of dimension
$2s+1$ if $\theta =0$. The carrier space of the representation
$(\theta,s)$, $\theta \neq 0$ is given by
 \begin{equation}
\stackrel{(\t,s)}{\cV}=\le\{\f\in\cL^2(\SU(2),\C) | \forall \xi \in
[-2\pi,2\pi], \f(xh_{\xi})=e^{is\xi}\f(x)\ri\},
\end{equation}
where $ h_{\xi} \in U(1)$ is a diagonal group element defined in
appendix
\ref{app:su2}.
This carrier space admits a decomposition in terms of
a direct sum of $SU(2)$ invariant subspaces
\begin{equation}
\wmod{\t,s}=\bigoplus_{I-s \in \N} \wmod{\t,s}_I.
\end{equation}
A  basis of $\wmod{\t,s}_I$  is given by the Wigner
functions (matrix elements of representations)
\begin{equation}
\{D^I_{m\, s}(x) \, \forall I \geq n,\ -I\leq m \leq I\}.
\end{equation}
The carrier space of the representation $(0,s)$ is the usual $\SU(2)$
spin $s$
representation and can be described as a functional space by
\begin{equation}
\stackrel{(0,j)}{\cV}=\le\{\f\in\cL^2(\SU(2),V^s) | \forall h \in
\SU(2),
\f(xh)=D^s(h^{-1})\f(x)\ri\}.
\end{equation}

The representation conjugate to $(\theta,s)$ is $(-\theta,-s)$.
The R-matrix is acting on a tensor product of representation
 by
\be
(\wrep{\t_1,s_1}\ox\wrep{\theta_2,s_2})({R}) \phi(x)\otimes \psi(y) =
\phi(x)\otimes \psi(xh_{\theta_1}^{-1}x^{-1}y).
\ee
and the ribbon element is acting on $\wmod{\t,s}$ by multiplication
with $ e^{is\theta}$.
The braiding element is ${\cal{R}}= \sigma \cdot R$ with $\sigma$ the
permutation:
$\sigma(u\otimes v) = v\otimes u$.

Among all these representations, we will
have a particular interest in the so-called \textit{simple
representations}. They are the representations of the form
$(\t,0)$ and $(0,j)$. The characteristic property of these
representations comes from the fact that the ribbon element, which
is central, is acting trivially on simple representations.
It is important to note that if $\theta_1=0$ the $R$-matrix
is acting as the identity operator.
In particular  the representation of the braid group
induced by this $R$ matrix is equivalent to the permutation group
if we restrict to the representations $(0,j)$.

With this data we can easily construct the Reshetikhin-Turaev
evaluation
of $D(\SU(2))$ colored framed tangles.
The key property that we need from the $R$ matrix is the fact that
The image of   $\wmod{\t_1,s_1}_I\otimes \wmod{\t_2,s_2}_J$ under the
action
of $R$ is contained in $(\oplus_L \wmod{\t_1,s_1}_L)\otimes
\wmod{\t_2,s_2}_J$
where the sum is over a finite number of $L$, therefore no infinities
arise in
open tangle evaluation.

An immediate and serious problem arise when we try to close the
tangles to construct link invariant: $\cD(\SU(2))$ is a non compact
quantum group and some of its unitary representations are infinite
dimensional. In this case taking the trace is not allowed unless
the tangle operator is proven to be trace class. In the case of a
knot this can be easily circumvented \cite{martins}. It can be
easily proven that the tangle of a colored  knot $K_\rho$ is a
central element and therefore acts on $V_{\rho_{C_b}}$ as a scalar
and the value of this scalar is taken to be the knot
invariant. The case of a general link is far more complicated and
there is no known way to construct link invariants associated with
infinite dimensional representations. We don't know if it is
possible to define a Reshetikhin-Turaev evaluation of any link
using a non compact quantum group but we will show that we can
define such an evaluation for colored Chain mail links. But first,
we  need the following notion of \textit{reduced trace}.
\begin{defin}[Diagonal endomorphism and reduced
trace]\label{def:redtrace}  The trace of an endomorphism $\cE$ of
$\wmod{\t,0}$ is the sum of the trace over each of the subspaces
$\wmod{\t,0}_I$. An endomorphism $\cE$ is said to be
\textbf{diagonal} when its trace over each $\wmod{\t,0}_I$ is equal
to $(2I+1) T_{\cE}$ where $T_{\cE}$ is independent of $I$. We then
define $T_{\cE}$ as the \textbf{reduced trace} of $\cE$ over
$\wmod{\t,0}$.
\end{defin}

\subsection{Definition of the chain-mail invariant for $\cD(\SU(2))$}

In this section, we define the chain-mail invariant, using
$\cD(\SU(2))$, associated to a manifold $\cM$. This construction
depends also on the choice of a triangulation $\D$, and maximal
trees $\cT$ and $\cT^*$. We will not directly prove  that the
result is independent of these ingredients but we are going to show
that
it is equivalent with the
previously defined Ponzano-Regge invariant and we will refer to it as
chain-mail invariant.

\begin{defin}[Amplitude for a colored link]
Let us label each $C$-component $e$ of $L_{\D,\cT,\cT^*}$ by a
simple representation $(0,j_e)$ of $\cD(\SU(2))$ and each
$C^*$-component $e^*$ by a simple representation $(\t_{e^*},0)$.
We obtain this way the colored oriented framed link $\ljt$.

We have the following key property:
There exists a presentation of this link as a closure
of a framed tangle such that the Reshetikhin-Turaev evaluation
of this tangle gives an endomorphism of
$\otimes_{e^*}\wmod{\t_{e^*},0}$
which is diagonal for each factor $\wmod{\t_{e^*},0}$.

We define the evaluation $$\eval[\ljt]$$ of this colored link
to be the reduced trace of this tangle evaluation.
Moreover since the representation are simple $\eval[\ljt]$
does not depend on the framing of the link.
\end{defin}

The fact that the endomorphisms of $\wmod{\t,0}$ are all diagonal
and that the notion of reduced trace apply will be proved later,
when we will give an explicit computation of this quantity (see
proposition \ref{prop:simpletrace}). It should be noted that due to
this notion of reduced trace, we are able for the first time to
define an invariant using infinite dimensional representations.
However we do not claim that such an invariant exist for every
link. This makes sense only because we only have to trace
\textit{diagonal} endomorphism in the evaluation process. We
conjecture that this is a property that holds for every link
having the properties 1,2,3 listed above for the chain-mail link,
namely the fact that it is made of unknot component that can be
separated in two types, such that components of the same type are
not linked together and such that  one type of component
is colored by finite dimensional representations.

\begin{defin}[Chain-mail invariant]
The chain-mail invariant is defined as the sum over
representations labels (colors), of the evaluation of the colored
link.
\begin{equation}\label{eqn:CMdef}
CM(\cM,\D,\cT,\cT^*)=\le(\prod_{e\not\in\cT} \sum_{j_e}
d_{j_e}\ri) \le(\prod_{e^*\not\in\cT^*} \int_{H/W}d \t_{e^*}
\D^2(\t_{e^*})\ri)\ \ \eval[L_{\D,\cT,\cT^*}(\t_{e^*},j_e)]
\end{equation}
\end{defin}

\subsection{Main theorem}

So far we did not prove that $CM(\cM,\D,\cT,\cT^*)$ is independent
of all the ingredients and is actually an invariant of $\cM$. This
is proved by the fact that it is equal to the Ponzano-Regge
invariant. This is the main result of this paper
which gives rise to the following theorem
\begin{theo}\label{theo:eqampli}\label{theo:CMPR}
The Ponzano-Regge amplitude we defined for a colored triangulation
is equal to the Reshetikhin-Turaev evaluation of the colored
chain-mail link
\begin{equation}\label{eqn:eqamplit}
\zjt=\eval(\ljt)
\end{equation}
It then follows from (\ref{eqn:PRsumdef}) and (\ref{eqn:CMdef}) that
\begin{equation}
CM(\cM,\D,\cT,\cT^*)=PR(\cM_d).
\end{equation}
$CM(\cM,\D,\cT,\cT^*)$ is thus an invariant of $\cM$ that can be
denoted $CM(\cM)$. It is equal to the gauge fixed Ponzano-Regge
invariant.
\end{theo}

To prove the theorem, we first
rewrite the LHS of (\ref{eqn:eqamplit}) in a convenient way,
then we rewrite the RHS as the evaluation of a product
of tangles that can be easily computed. In particular
 we prove that the endomorphism to be
traced are diagonal.

\begin{prop}
Let us consider the following set of indices
\begin{equation}
P=\le\{(e,e^*),\ s.t.\ e\subset e^*\ri\}
\end{equation}
and we denote by $I=(e,e^*)$ an element of $P$. This is a finite
set. For a choice of $I=(e,e^*)$, we denote $j_I=j_e$,
$\t_I=\t_{e^*}$, $x_I=x_{e^*}$ and $\e(I)=\e(e,e^*)$. For
$I=(e,e^*)$, we denote by $s(I)$ the index $(e,\te^*)$ where
$\te^*$ is the successor of $e^*$ inside the dual face $f^*\sim
e$, according to the cyclic order induced by orientation of $e$.
The amplitude $\zjt$ can be written as
\begin{equation}\label{eqn:zjtbis}
\zjt=\int_{G/H} \prod_{e^*}dx_{e^*}\  \prod_I
D^{j_I}_{a_Ib_I}(x_Ih_{\t_I}^{\e(I)}x_I^{-1}) \ \times \prod_I
\d_{b_I,a_{s(I)}}
\end{equation}
\end{prop}
\proof{This comes from splitting the characters of definition
\ref{def:zjt} into matrix elements of representations of $\SU(2)$.}

\vspace{1ex}

\begin{prop}\label{prop:simpletrace}
Let us pick up an arbitrary order of the dual edges $e^*$. The
link $L_{\D,\cT,\cT^*}$ can be written as the closure of the
following product of colored tangles
\begin{center}
\psfrag{T}{$T$}\psfrag{B}{$B$}\includegraphics[height=2cm]{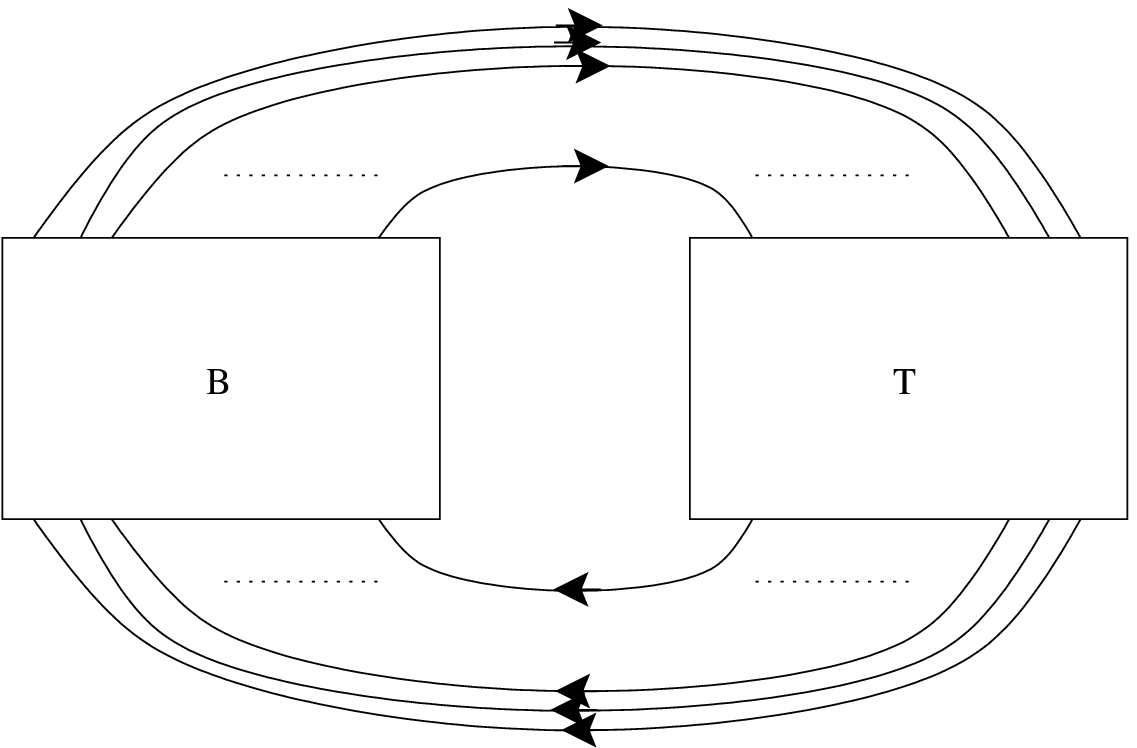}

\end{center}
where
\begin{itemize}
\item $T$ is a colored tangle
\begin{equation}
T : \bigotimes_I V^{j_I} \to \bigotimes_I V^{j_I},\ \ \ \ \
T=\bigotimes_{e^*} T_{e^*}
\end{equation}
with
\begin{equation}
T_{e^*}:\bigotimes_{e\subset e^*} V^{j_e} \to
\bigotimes_{e\subset e^*} V^{j_e}, \ \ \ \ \ \ \ T_{e^*} =\ \ \
\figeq{\psfrag{c}{$\t$}\psfrag{b1}{$j_1$}\psfrag{b2}{$j_2$}\psfrag{b3}{$j_3$}\psfrag{bn}{$\!\!\!j_{N-1}$}\psfrag{bn1}{$\
\ j_N$}\includegraphics[width=4cm]{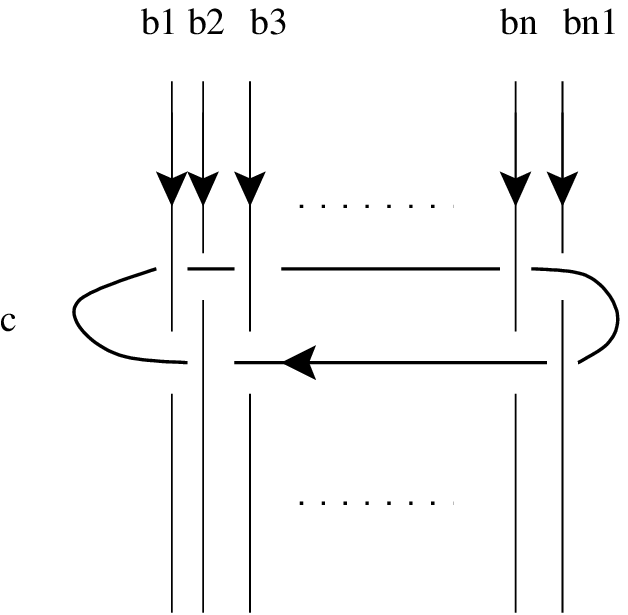}}%
\end{equation}
The relative crossing between the lines depending on the relative
orientation of $e^*$ and $e$.
$T_{e^*}$ is the closure of a colored braid $B_{e^*}$. The
Reshetikhin-Turaev evaluation
of $B_{e^*}$ is a diagonal endomorphism of
$\wmod{\t,0} \otimes_{e\subset e^*} V^{j_e}$, i-e
$tr_{\wmod{\t,0}_I}(B_{e^*})/(2I+1)$
is independent of $I$ and defines the evaluation of
$T_{e^*}$ as an endomorphism of $ \otimes_{e\subset e^*} V^{j_e}$.
\item $B$ is a colored braid
\begin{equation}
B : \bigotimes_I V^{j_I} \to \bigotimes_I
V^{j_{s(I)}}.
\end{equation}
\end{itemize}
\end{prop}
{\it Proof:}

{The link $L_\D$ is such that the $C^*$-components carrying
representations $\t^*$ are unlinked together. We can thus isolate
the $C^*$-components and put them according to the arbitrary order
we choose for the $e^*$. We move the $C$-components in order that
they all cross the $C^*$-components in the same direction. We then
obtain the tangle $T$ as a product of $T_{e^*}$. Now $T$ is closed
by a braid that permutes the strands of the $C$-components from
$\bigotimes_I V^{j_I}$ to $\bigotimes_I V^{j_{s(I)}}$.
Since the action of the $R$-matrix is trivial
on representations $(0,j)$ We don't
need to specify the detailed structure of this braid since
 it is only its permutation effect that will matters.
 Thus the
Reshetikhin-Turaev evaluation of the colored braid $B$ is given by
\begin{equation}
\prod_I \d_{b_I,a_{s(I)}}.
\end{equation}
$T_{e^*}$ is given in appendix \ref{app:braidings}. It is proven there
that  the trace over the representation space
$\wmod{\t,0}$ involves a diagonal endomorphism. The evaluation can
therefore be understood in terms of
reduced trace and is equal to
\begin{equation}\label{linkCj}
\figeq{\psfrag{c}{$\t_{e^*}$}\psfrag{b1}{$j_{1}$}
\psfrag{b2}{$j_2$}\psfrag{b3}{$j_3$}\psfrag{bn}{$\!\!\!j_{N-1}$}
\psfrag{bn1}{$\ \ j_N$}
\includegraphics[width=4cm]{braidn2.eps}} = \int_{G/H}
dx_{e^*} \bigotimes_{e\subset e^*}
 D^{j_e}(x_{e^*}h_{\t_{e^*}}^{\e(e,e^*)}x_{e^*}^{-1})
\end{equation}
 }$\Box$

The combination of these results then show that the evaluation of
the colored link $\ljt$ gives the amplitude $\zjt$ given by
(\ref{eqn:zjtbis}) and proves the theorem.

\subsection{Properties of the chain-mail construction}

In this section we investigate some properties of the chain-mail
construction that could be used to give a direct proof of the
invariance properties of the chain mail evaluation. To make
contact with the chain mail notations of Roberts \cite{sktv}, one
also introduce the representation $\O$ to color $E$-type
components and the representation $\O^*$ to color $E^*$-type
components. They correspond to formal linear combinations of
the $(0,j)$ and $(\t,0)$ color
\begin{eqnarray}
\O&=&\sum_j d_j\ (0,j) \\ \O^* &=&\int_H \frac{d\t}{\pi} \D(\t)^2\ (\t,0)
\end{eqnarray}
With this definitions, we have the following result
\begin{prop}
The chain-mail invariant $CM(\cM)$ is equal to the evaluation of
the link $L_{\D,\cT,\cT^*}$ where we colored all the
$C$-components by the representation $\O$ and all the
$C^*$-components by the representation $\O^*$.
\end{prop}

\subsection{Sliding, killing and fusion properties}
We have already used some evaluations properties of $\cD(\SU(2))$ to prove
the equality of amplitudes in theorem \ref{theo:eqampli}.
Evaluation of elementary tangles are collected and proven in
appendix \ref{app:braidings}.

Particular features of Roberts chain-mail calculus are the
so-called sliding and killing properties satisfied by colored
link, and the fusion identities. They are exposed here and proved
in appendix \ref{app:braidings}.

\begin{prop}
The sliding properties are
\begin{center}
\psfrag{Oj}{$\O$} \psfrag{Ot}{$\O^*$}
\psfrag{j}{$j$}\psfrag{t}{$\t$}
\includegraphics[width=8cm]{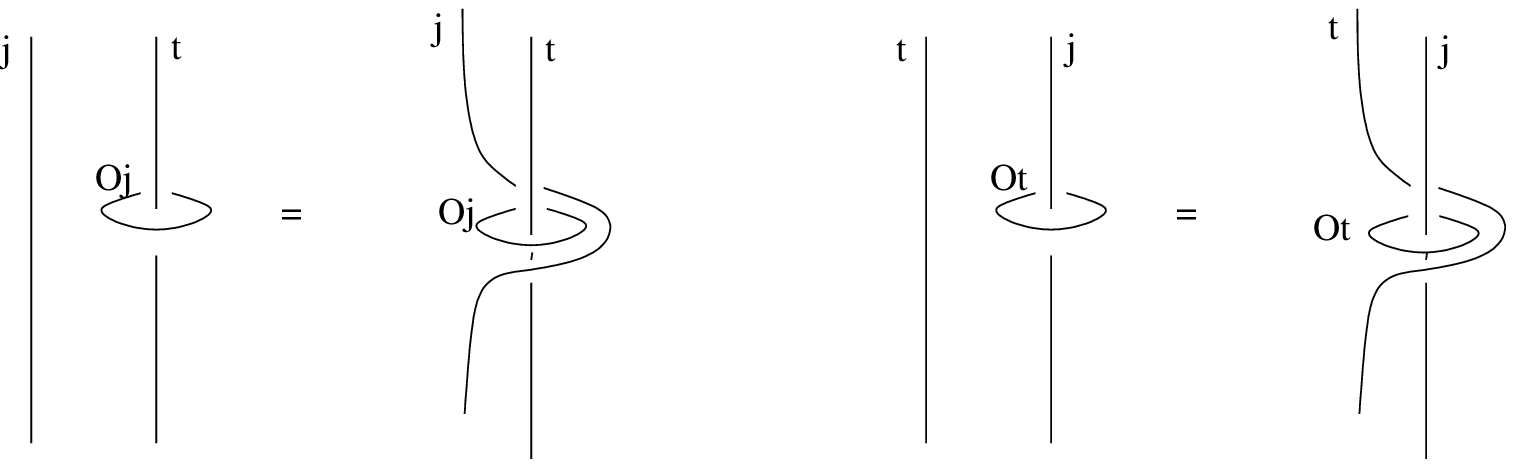}
\end{center}

The killing properties are
\begin{center}
\psfrag{Oj}{$\O$} \psfrag{Ot}{$\O^*$}
\psfrag{j}{$j$}\psfrag{t}{$\t$}
\psfrag{dj}{$\d_{j,0}$}\psfrag{dt}{$\d(h_{\t})$}
\includegraphics[width=8cm]{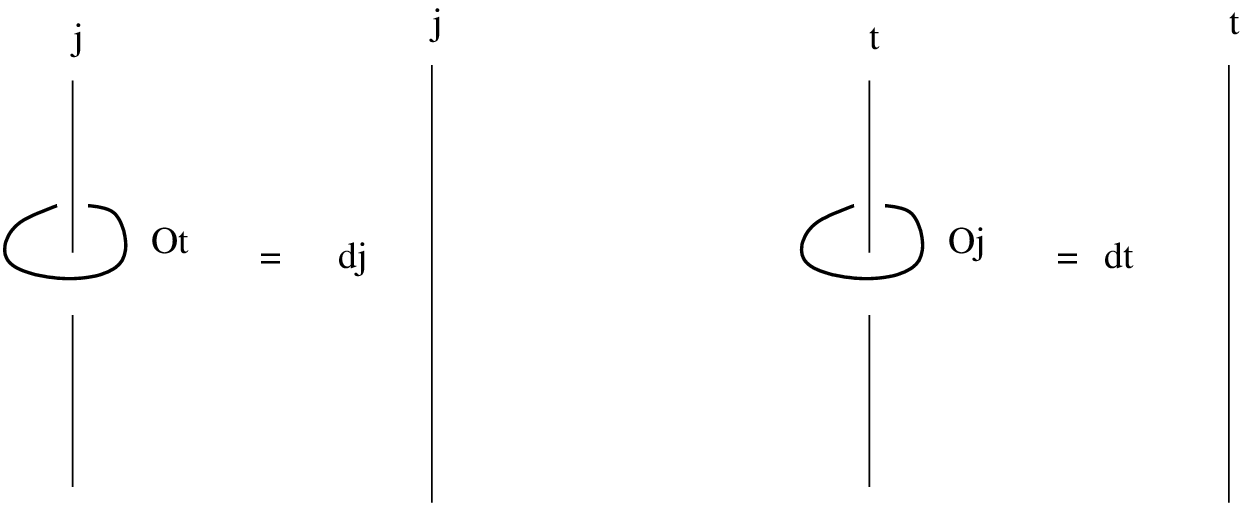}
\end{center}

The fusion properties are
\begin{equation}
\psfrag{t}{$\t$}\psfrag{j1}{$j_1$}\psfrag{j2}{$j_2$}
\psfrag{r1}{$j_1$} \psfrag{r2}{$j_2$} \psfrag{r3}{$j_3$}
\psfrag{R}{$\t$}
\figeq{\includegraphics[height=3cm]{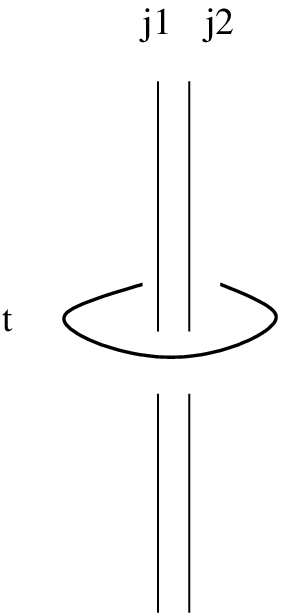}}=\sum_{j_3}d_{j_3}
\figeq{\includegraphics[height=3cm]{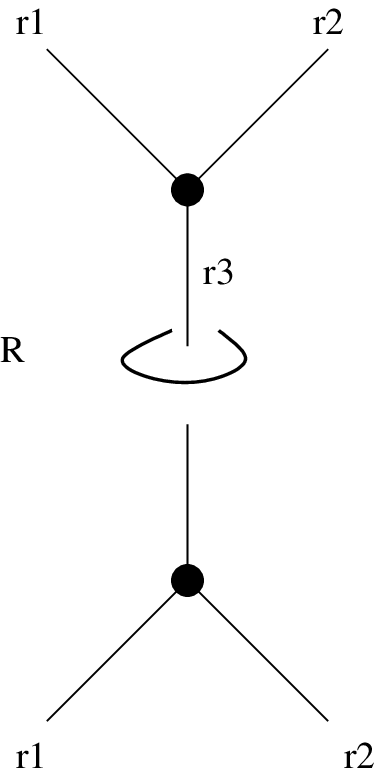}},\ \hspace{2cm}
{\psfrag{t}{$j$}\psfrag{j1}{$\t_1$}\psfrag{j2}{$\t_2$}
\figeq{\includegraphics[height=3cm]{braid2.eps}}}= \int d\v(\t_3) \sum_{s_3}
\psfrag{r1}{$(\t_1,0)$}\psfrag{r2}{$(\t_2,0)$}
\psfrag{r3}{$(\t_3,s_3)$}\psfrag{R}{$(0,j)$}
\figeq{\includegraphics[height=3cm]{fusion21.eps}}
\end{equation}
where \begin{equation} \psfrag{r1}{$j_1$}
\psfrag{r2}{$j_2$}\psfrag{r3}{$j_3$}\psfrag{x1}{}\psfrag{x2}{}\psfrag{x3}{}
\figeq{\includegraphics[width=2cm]{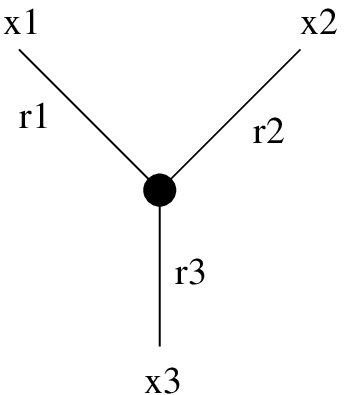}}, \ \hspace{2cm}
{\psfrag{r1}{$\t_1$}\psfrag{r2}{$\t_2$}\psfrag{r3}{$(\t_3,s_3)$}
\psfrag{x1}{}\psfrag{x2}{}\psfrag{x3}{}
\figeq{\includegraphics[width=2cm]{clebsh.eps}}}
\end{equation}
respectively denotes the Clebsh-Gordan coefficient of the
decomposition of the tensor product of representation
$(0,j_1)\ox(0,j_2)$ into $(0,j_3)$ (equal to the one of $\SU(2)$),
and the Clebsh-Gordan coefficient of the tensor product of
representations $(\t_1,0)\ox(\t_2,0)$ into $(\t_3,s_3)$
\footnote{Note that contrary to $(0,j)$ case, the tensor product
of two $(\t,0)$ representations does not decompose only on simple
representations.}, $d\v(\t_3)=\D(\theta_3) d\theta_3=
\sin(\theta_3/2)^2 d\t_3$ being the Plancherel measure (explicit
expression for the coefficient and the measure are given in
appendix \ref{app:braidings}).
\end{prop}


\section{Insertion of Wilson lines and Hilbert space}\label{sec:obs}

In this section we show that we can extend
 the construction of the Ponzano-Regge invariant in order to define
insertion of Wilson loops, and insertion of what we will call
\textit{particle graphs}. For each case we present the additional
data we need and the type of invariant we want to define. Then we
present  the
chain-mail version of the corresponding invariant:
 A non-compact topological field theory\footnote{
 We call non-compact a topological field theory
 where we relax the axiom of having only  finite dimensional Hilbert
space.}

We now consider the inclusion of Wilson loops (or rather Wilson
graphs) into the model. In the continuum, Wilson loops are defined
as the trace, in a certain representation, of the holonomy of the
connection along a closed loop. This concept is generalized by the
notion of spin-networks.
\begin{defin}[Spin-networks]
Let us consider an abstract oriented graph $\G$ with edges $e$
colored by $\SU(2)$ representations $j_e$ and vertices colored by
invariant tensors $\i_v$
that intertwines the representations of
corresponding edges according to their orientations.
More precisely $\i_v\in (\otimes_{e\supset v} V_{j_e}^{\e{v,e}})$
where
$\e(e,v) =+1$ if $e$ is ingoing at $v$ and $\e(e,v) =+1$ if $e$ is
outgoing
and $V^{+1}\equiv V, V^{+1}\equiv V^*$.
The
spin-network based on this colored graph is the function of $|e|$
group elements
\begin{equation}
\F_{\G,j_e,\i_v} : \{g_e\} \to \bra \bigotimes_e  D^{j_e}(g_e)
|\bigotimes_v \i_v\ket.
\end{equation}
Where the pairing is the natural pairing
between
$\bigotimes_{e}(V_{j_e}\otimes V_{j_e}^*)$ and
$\bigotimes_{v}(\otimes_{e\supset v} V_{j_e}^{\e(v,e)})$.
It should be noted that in the case of a trivalent graph $\G$,
intertwiners do not need to be specified since it exists only zero
or one normalized intertwiner for three representations of
$\SU(2)$.
\end{defin}
The spin-networks based on a graph $\G$ are functions of $|e|$
group elements of $\SU(2)$, and form an orthogonal
basis of the  Hilbert space
 $\cH_\G  \equiv \cL^2(\SU(2)^{|e|}/\SU(2)^{|v|},d^{|e|}g)$,
 where $d^{|e|}g$ is the product of
normalized Haar measure.
And the action of $\SU(2)^{|v|}$ on $\SU(2)^{|e|}$ is given by
$g_e\to k_{t_e}g_e k_{s_e}$, with $s_e,t_e$ the starting and terminal
vertex of $e$.

\subsection{Ponzano-Regge invariant with Wilson graph}

We define a colored Wilson graph $\G_C=(\G,j_{e^*},\i_{v^*})$
as subgraph of the dual 1-complex of $\D^*$, with edges $e^*$
colored by representations $j_{e^*}$ of $\SU(2)$, and vertices
$v^*$ colored by intertwiners $\i_{v^*}$. To define the
Ponzano-Regge invariant with insertion of a Wilson graph, we take
the same prescription as before for the maximal trees $\cT$ and
$\cT^*$.

\begin{defin}
Consider a decorated triangulation $\D_d$ of $\cM$, with
$d=(or,st,T,T^*)$
We define
 $\G_C\equiv (\G,j_{e^*},\i_{v^*})$ to be a colored subgraph of
$\D^*$
 if $\G\in \D^*_1$ is a subgraph of the dual 1-complex of $\D^*$,
with edges $e^*$
colored by representations $j_{e^*}$ of $\SU(2)$, and vertices
$v^*$ colored by intertwiners $\i_{v^*}$.
We also
define
\begin{equation}
PR[\cM,\D_d,\G_C]=\le(\prod_{e^*\in \D^*} \int_G
dg_{e^*}\ri)\prod_{e^*\in\cT^*} \d(g_{e^*}) \ \prod_{e\not\in\cT}
\d(G_e)\ \F_{\G_C}(\{g_{e^*}\})
\end{equation}
\end{defin}

In section \ref{def} we have seen that the data $\D_d$ defines a
measure $\mu$ (\ref{de},\ref{locaint}) on the space ${\cal A}_F$
of $SU(2)$ flat connection. Given a colored graph we can construct
a spin network functional $\F_{\G_C}$ on ${\cal A}_F$ and
$PR[\cM,\D_d,\G_C]$ then compute the integral $\mu(\F_{\G_C})$. It
is easy to prove using techniques of appendix \ref{lemma1} that
$PR[\cM,\D_d,\G_C]$ depends only on the homology class of $\G$. A
more challenging question is wether one can generalize theorem 1
and show that not only the total volume of ${\cal A}_F$ is
invariant under the change of decorated triangulation but the
measure $\mu$ itself. We will not answer this question here but
focus on the relation of this definition with a chain mail
evaluation.

\begin{defin}\label{def:zjtg}
Lets consider a decorated triangulation $\D_d$ of $\cM$, colored by
$j_e$ for each edge
$e\not\in T$ and $\t_{e^*}$ for dual edge $e^*\notin T^*$, together
with $\G_C$ a colored subgraph of $\D^*$
We associate to these data the amplitude
\begin{equation}\label{eqn:zjtdefg}
Z_{\D_{d},\G_C}(\{j_e\},\{\t_{e^*}\})=
\int_{G/H} \prod_{e^*}  dx_{e^*}
\prod_{e} \X_{j_e}(\prodor{e^*\subset e}
x_{e^*}h^{\e(e^*,e)}_{\t_{e^*}}x_{e^*}^{-1})
\F_{\G_C}(\{x_{e^*}h_{\t_{e^*}}x_{e^*}^{-1}\}),
\end{equation}
where $\X_{j_e}$ is the character of the spin $j_e$
representation, and $h_{\t_{e^*}}\in SU(2)$, is the representative
element
of $\t_{e^*} \in U(1)$ given in appendix \ref{app:su2}.
\end{defin}

Given the data $\D, \G_C$ we can construct a graph
$L_{{\D,\G}} = L_{\D}\cup \G$
which consist of the chain mail link $L_{\D}$  knotted with $\G$ as
follows:
As described previously, the chain mail link $L_{\D}$ is obtained,
for
a closed manifold, from $\D$ by considering the handlebodies
$H,H^{*}$
which are the thickening of the one skeleton of $\D$ and $\D^{*}$.
In the boundary surface of $H^{*}$ we draw the collection $C_{e^{*}}$
of $H^{*}$'s meridians  and the  collection $C_{e^{*}}$
of $H$'s meridians, moreover
in the core of $H^{*}$ we put the graph $\G$.
The component $C_{e}$ and $C_{e^{*}}$ intersects each other on
$\partial H^{*}$.
In order to obtain a link we push the $C_{e}$ components inside
$H^{*}$  in such a way that these components are not link with
the graph $\G$ in the core of $H^{*}$.
The resulting object is $L_{\D,\G}$.

We label each $C$-component $e$ of $L_{\D,\G}$ by a
simple representation $(0,j_e)$ of $\cD(\SU(2))$ and each
$C^*$-component $e^*$ by a simple representation $(\t_{e^*},0)$ and
we color the graph $\G$ by $C\equiv ((0,j_{e^*}),\i_{v^*})$.
We obtain in this way a  graph colored by $\cD(\SU(2))$
representations and intertwiners denoted
$L_{\D,\G}(\{j_{e}\},\{\t_{e^{*}}\},C)$.

\begin{prop}\label{wilampli}
The  Wilson graph amplitude $Z_{\D_{d},\G_C}(\{j_e\},\{\t_{e^*}\})$
amplitude is equal to the Reshetikhin-Turaev evaluation of
$L_{\D,\G}(\{j_{e}\},\{\t_{e^{*}}\},C)$.
\end{prop}
{This proof is similar to the proof of theorem \ref{theo:eqampli}.
It was shown that the evaluation of the chain mail link gives an
integral over $\{x_{e^{*}}\}_{e^{*}\in \D^*}, x_{e^{*}}\in G/H$ of
$ \prod_{e} \X_{j_e}(\prodor{e^*\subset e}
x_{e^*}h^{\e(e^*,e)}_{\t_{e^*}}x_{e^*}^{-1})$. Now each edge
$e^{*}¥$ of the graph $\G$ colored by $j_{e^{*}}^{\G}$ is linked
with a component $C_{e^{*}}$ colored by $\t_{e^{*}}$. As shown in
(\ref{linkCj},\ref{Testar1})  the Reshetikhin-Turaev evaluation of
such a linking produces in the integrand an additional term
$D^j(x_{e^{*}}h_{\t_{e^{*}}}x_{e^{*}}^{-1})$, the contraction of
this terms with the vertex intertwiners  reproduces
$Z_{\D_{d},\G_C}(\{j_e\},\{\t_{e^*}\})$.}

\subsection{Ponzano-Regge invariant with Particle graph}

In reference \cite{FLDSU2phys} a new class of Ponzano-Regge
observables allowing the insertion of interacting spinning
particles coupled to gravity have been considered. In this section
we  show that the expectation value of such observables can be
reproduced by a Reshetikhin-Turaev evaluation.

A particle graph, denoted $\da$ is  a subgraph of $\D^{*}$
submitted to a marking of bivalent vertices. Generically a
subgraph of $\D^{*}$ contains bivalent vertices and (true)
multivalent vertices. Each bivalent vertex $v^{*}$ belong to
several\footnote{A dual vertex is dual to a tetrahedra, dual faces
which meet in $v^{*}$ are dual to the edges of this tetrahedra,
there are therefore 6 such dual faces} dual faces, only one of
which do not intersect $\da$ along an edge. Since dual faces are in one to one
correspondence with edges of the triangulation, this means that we
can associated a unique edge of $\D$ to  a bivalent vertices of
$\da$.

A marking of $\da $ is a choice of bivalent vertices of $\da$ such
that the corresponding set of edges form a subgraph of $\D$,
denoted $\tilde{\da}$, which is isotopic to $\da$. The graph $\da
$ provides $\tilde{\da} $ with a framing.
We restrict the graph and the marking to the case where
two different marked vertex corresponds to different edges of
$\tilde{\da} $.

A coloring $C$ of $\da$ is a choice of $SU(2)$ representation
$I_{e^*}$ for each edge $e^{*}$ of $\da$, a choice of $SU(2)$
intertwiner $i_{v}$ for each unmarked vertex  and a choice of a
$\cD(SU(2)) $ representation $(m_{e}¥,s_{e}¥)$ to each marked bivalent vertex $e\in
\tilde{\da}$. \footnote{Physically, the subgraph $\tilde{\da}$ of
$\D$ associated with the particle graph $\da$ consist of particle
worldlines, and $(m,s)$ denote the mass and spin of the particles
(see \cite{FLDSU2phys})}.

\begin{defin}[Particle graph functional]
Given a decorated particle graph $\da_{C}$, we construct a functional
depending on $|e^{*}|+|e|$ group elements where
$|e^{*}|$ is the
number of edges of $\da$ and $|e|$ the number of marked vertices (or
edge of $\tilde{\da}$):
\begin{equation}
\Pi_{\da_{C}} : (\{g_{e^{*}}¥\}, \{u_{e}\})¥ \to \bra \bigotimes_{e^*}
 D^{I_{e^{*}}¥}(g_{e^*}) |\bigotimes_e \Pi^s_{I_{s_{e}I_{t_{e}}}}(u_{e})
\bigotimes_{v^*} \i_{v^*}\ket.
    \end{equation}
Where ${s_{e}},{t_{e}}$ are the two edges of $\da$ meeting at the
marked
vertex $e$ and
\begin{equation}\label{defpi}
    \Pi^s_{IJ}(u)=\sqrt{d_I d_J} D^{I}(u^{-1})|s,I\ket \bra J,s| D^{J}(u) \in
    V_{I}\otimes V_{J}^{*}
\end{equation}
 with $|s,I\ket $ the normalized vector of spin $s$ in $V_{I}$.
The pairing in (\ref{defpi}) is the natural pairing
 between
 $\bigotimes_{e^{*}}(V_{j_{e^*}}\otimes V_{j_{e^{*}}}^*)$ and
 $\bigotimes_{e}(V_{I_{{s_{e}}}¥}\otimes V_{t_{e}}^{*})
 \bigotimes_{v}(\otimes_{e^{*}¥\supset v} V_{j_e}^{\e(v,e)})$.
\end{defin}

Given these data we can construct the following amplitude
\begin{eqnarray}
Z_{\D_{d},\da_{C}}(\{j_e\},\{\t_{e^*}\})&=&
\int_{G/H} \prod_{e^*}  dx_{e^*}  \int_{G}\prod_{e\in \tilde{\da}}du_{e}
\prod_{e\notin \da} \X_{j_e}(\prodor{e^*\subset e}x_{e^*}h^{\e(e^*,e)}_{\t_{e^*}}x_{e^*}^{-1})\times \\
& & \prod_{e\in \da} \X_{j_e}(\prodor{e^*\subset e}x_{e^*}h^{\e(e^*,e)}_{\t_{e^*}}x_{e^*}^{-1}
u_{e}h_{m_e}u_{e}^{-1})
\Pi_{\da_{C}}(\{x_{e^*}h_{\t_{e^*}}x_{e^*}^{-1}\}, \{u_{e}\}).
\end{eqnarray}

\begin{figure}
\psfrag{C}{$C_e$} \psfrag{C^*}{$C^*_{e^*}$}
\psfrag{G}{$\da$}\psfrag{H}{$H^*$}
\includegraphics[width=6cm]{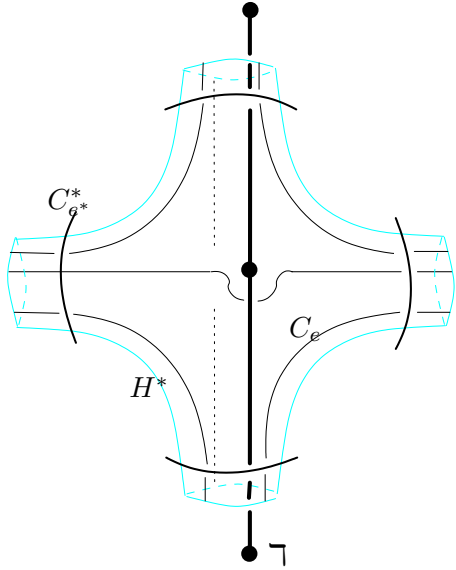}
\caption{The Chain mail around a marked vertex of $\da$}
\label{chainvertex}
\end{figure}

Given a particle graph we can construct a graph $L_{\D,\da}$ which
is the union of the chain mail $ L_{\D}$ with $\da$. As before in
order to obtain this graph we consider the Handlebody $H^{*}$, on
the boundary of which we draw the meridians of $H$ and $H^{*}$ and
we put $\da $ in the core of $H^{*}$. We push the $C_{e}$
components slightly inside $H^{*}$or the $C^*_{e^*}$ components slightly
outside $H^*$. Moreover, if $e$ correspond to
an edge of $\tilde{\da}$ we link the part of $C_{e}$ lying between
$C_{s_{e}}^* $ and $C_{t_{e}}^*$ with $\da$ by letting $C_{e}$ go
once around the core of $H^{*}$ (see figure \ref{chainvertex}).

We label each $C$-component $e$ of $L_{\D,\G}$ by a
simple representation $(0,j_e)$ of $\cD(\SU(2))$ and each
$C^*$-component $e^*$ by a simple representation $(\t_{e^*},0)$ and
we color the edges of  $\da$ by $ (0,j_{e^*})$ the
vertices by $\i_{v^*}$ and the marked bivalent vertices
by $\cD(\SU(2)) $ representations $(m_{e},j_{e})$.
We obtain in this way a  graph colored by $\cD(\SU(2))$
representations and intertwiners denoted
$L_{\D,\da}(\{j_{e}\},\{\t_{e^{*}}\},C)$.

\begin{prop}\label{wilampli2}
The  Wilson graph amplitude $Z_{\D_{d},\da_C}(\{j_e\},\{\t_{e^*}\})$
amplitude is equal to the Reshetikhin-Turaev evaluation of
$L_{\D,\da}(\{j_{e}\},\{\t_{e^{*}}\},C)$.
\end{prop}
{This proof is similar to the proof of the previous proposition
and the theorem \ref{theo:eqampli}. For  edges $e$ which do not belong to
$\tilde{\da}$
the evaluation of the chain
mail link gives an integral over $\{x_{e^{*}}\}_{e^{*}\in \D^*_1}, x_{e^{*}}\in
G/H$ of $ \prod_{e} \X_{j_e}(\prodor{e^*\subset e}
x_{e^*}h^{\e(e^*,e)}_{\t_{e^*}}x_{e^*}^{-1})$.
For edges $e$ which belong to
$\tilde{\da}$ there is an additional linking of
$C_e$ with an edge of $\da$ colored by $(m_{e},s_{e})$.
By construction $C_e$ is linked with only one edge of $\da$.
As shown in proposition
\ref{prop:Testar2} such a linking insert in the chain mail  an integral over $u_{e}$,
an element $u_eh_{m_e}u_e^{-1}$ in the trace over the spin $j$ representation,
giving a factor
$\X_{j_e}(\prodor{e^*\subset e} x_{e^*}h^{\e(e^*,e)}_{\t_{e^*}}x_{e^*}^{-1}u_eh_{m_e}u_e^{-1})$.
This linking also insert in the
evaluation of
$L_{\D,\da}(\{j_{e}\},\{\t_{e^{*}}\},C)$
a factor $\Pi^s_{I_{s_{e}I_{t_{e}}}}(u_{e})$.
Eventually, each edge
$e^{*}¥$ of the graph $\da$ colored by $j_{e^{*}}^{\G}$ is linked
with a component $C_{e^{*}}$ colored by $\t_{e^{*}}$. As shown in
(\ref{linkCj}, \ref{Testar1})  the Reshetikhin-Turaev evaluation
of such a linking produces in the integrand an additional term
$D^j(x_{e^{*}}h_{\t_{e^{*}}}x_{e^{*}}^{-1})$ which is contracted with
$\Pi$ and the graph intertwiners.
 Putting all these terms
together we recover $Z_{\D_{d},\da_C}(\{j_e\},\{\t_{e^*}\})$.}

\subsection{Gauge fixing operators and their evaluation}

If we start from the definition of the gauge fixed Ponzano-Regge invariant
(\ref{eqn:defregulPR}), we can perform only the Plancherel
decomposition given in prop. \ref{prop:PRstatesum} to get the
following expression
\begin{equation}\label{eqn:PRdelta}
PR(\cM)=\le(\prod_e \sum_{j_e} d_{j_e} \prod_{e^*} \int_{\SU(2)}
dg_{e^*}\ri)\le(\prod_{e\in\cT}\d_{j_e,0}\prod_{e^*\in\cT^*}\d(g_{e^*})\ri)
\le(\prod_e \X_{j_e}(\prodor{e^*\subset e}
g_{e^*}^{\e(e,e^*)})\ri).
\end{equation}
This makes apparent the fact that the regularization we proposed,
using maximal trees, can be performed by inserting the expression
\begin{equation}\label{eqn:gaugefix0}
\le(\prod_{e\in\cT}\d_{j_e,0}\prod_{e^*\in\cT^*}\d(g_{e^*})\ri),
\end{equation}
into the non-gauge fixed expression. This insertion of
$\d$-functions allows to avoid redundant sum and integral that
cause divergences. This expression is a particular case of the
notion of \textit{evaluation of an operator} that we define now.

\begin{defin}[Operator and its evaluation]
We define as an operator every function $\cO(j_e,g_{e^*})$ of the
labels $j_e$ and group elements $g_{e^*}$ for edges $e$ and dual
edges $e^*$ of $\D$. The evaluation of an operator is defined to
be
\begin{equation}
\le<\cO\ri>=\le(\prod_e \sum_{j_e} d_{j_e} \prod_{e^*}
\int_{\SU(2)} dg_{e^*}\ri)\ \ \cO(j_e,g_{e^*})\ \ \le(\prod_e
\X_{j_e}(\prodor{e^*\subset e} g_{e^*}^{\e(e,e^*)})\ri).
\end{equation}
\end{defin}

According to this definition, the gauge fixed expression
(\ref{eqn:PRdelta}) corresponds then to the evaluation of the
operator (\ref{eqn:gaugefix0}). This type of operator can be
generalized to the so-called \textit{gauge fixing operators}.

\begin{defin}[Gauge fixing operators]
For a choice of a graph $\G$ of $\D$, a dual graph $\G^*$, and an
assignment of labels $j_e^\G$ and $\t_{e^*}^{\G^*}$, we define
the gauge fixing operator
\begin{equation}
\cO_{\D,\G,\G^*,j_e^\G,\t_{e^*}^{\G^*}}(j_e,g_{e^*})=\prod_{e\in\G}
\frac{\d_{j_e,j_e^\G}}{d_{j_e^\G}^2}\prod_{e^*\in\G^*}
\d_{\t_{e^*}^{\G^*}}(g_{e^*}),
\end{equation}
where $\d_{\t_{e^*}^{\G^*}}$ is defined by
\begin{equation}
\int_G f(g)\d_{\t}(g)=\int_{G/H} du f(uh_\t u^{-1})
\end{equation}
projects $g_{e^*}$ onto the conjugacy class $\t_{e^*}^{\G^*}$ of
$\SU(2)$.
\end{defin}

\subsection{Examples of gauge fixing operators}

Among the possible gauge fixing operators, we notice the following

 As we have seen, the gauge fixed Ponzano-Regge invariant can be
rewritten
as the evaluation of the operator obtained by taking $\G=\cT$ with
$j_e^\G=0$ and $\G^*=\cT^*$ with $\t_{e^*}^{\G^*}=0$.
\begin{equation}
PR(\D,\cT,\cT^*)=\le<\cO_{\D,\cT  ,\cT  ^*,0,0}\ri>
\end{equation}
The meaning of the invariance of the choices of trees proved in
theorem \ref{theo:PRinv} is that this evaluation does not depend
on the choice of $\cT$ and $\cT^*$
\begin{equation}
\le<\cO_{\D,\cT_1,\cT_1^*,0,0}\ri>=\le<\cO_{\D,\cT_2,\cT_2^*,0,0}\ri>
\end{equation}
for every choices of $(\cT_1,\cT^*_1)$ and $(\cT_2,\cT^*_2)$.

 We
consider the operator where $\G$ is the 1-skeleton $\D_1$ of $\D$,
$\G^*$ its dual 1-skeleton $\D_1^*$. If we choose labels
$j_e^\G=0$ on $\cT$ and $\t_{e^*}^{\G^*}=0$ on $\cT^*$ and generic
labels $j_e^\G$ $\t_{e^*}^{\G^*}$ elsewhere, we get the amplitude
\begin{equation}
\le<\cO_{\D,\D_1,\D_1^*,j_e,\t_{e^*}}\ri>=Z_{\D,\cT,\cT^*}(\{j_e^\G\},\{\t_{e^*}^{\G^*}\}).
\end{equation}

 The notion of evaluation of operator leads us to the possibility
of modifying the definition (\ref{eqn:PRdelta}) by putting labels
fixing on the trees other than the zero. We consider the operator
$\cO_{\D,\cT,\cT^*,j_e^\cT,\t_{e^*}^{\cT^*}}$ instead of
$\cO_{\D,\cT  ,\cT  ^*,0,0}$. Its evaluation is given by the next
theorem.

\begin{theo}
We consider the gauge fixing operator
$\cO_{\D,\cT,\cT^*,j_e^\cT,\t_{e^*}^{\cT^*}}$ whose graphs are
maximal trees, and labels are generic labels $j_e^\cT$ and
$\t_{e^*}^{\cT^*}$.
\begin{itemize}
\item The evaluation of operator
$\le<\cO_{\D,\cT,\cT^*,j_e^\cT,\t_{e^*}^{\cT^*}}\ri>$ is equal to
the Reshetikhin-Turaev evaluation of the chain-mail link $L_\D$
colored by representations $\O$ and $\O^*$, except the components
on $\cT$ and $\cT^*$ that are colored by $j_e^\cT$ and
$\t_{e^*}^{\cT^*}$
\item We then have the following equality
\begin{equation}\label{eqn:BRSTcolor}
\le<\cO_{\D,\cT,\cT^*,j_e^\cT,\t_{e^*}^{\cT^*}}\ri> =
\le<\cO_{\D,\cT,\cT^*,0,0}\ri>= PR(\cM_d)
\end{equation}
\end{itemize}
\end{theo}

\proof  As described in appendix \ref{app:su2}, we can relate
$\d_{\t}(g)$ to a delta function on $H/W$, we get
\begin{equation}
\d_\t(h_\phi)=\frac{\pi}{\D^2(\t)}\d_\t(\phi)
\end{equation}
Let us rewrite
$\le<\cO_{\D,\cT,\cT^*,j_e^\cT,\t_{e^*}^{\cT^*}}\ri>$ in terms of
explicit formula, using the Weyl formula, we get
\begin{equation}
\le<\cO_{\D,\cT,\cT^*,j_e^\cT,\t_{e^*}^{\cT^*}}\ri>
=\le(\prod_{e\not\in\cT} \sum_{j_e} d_{j_e}
\prod_{e^*\not\in\cT^*}\frac{d\t_{e^*}}{\pi} \int_{H/W}
\D^2(\t_{e^*})\ri)\prod_{e^*}\int_{G/H} dx_{e^*} \prod_e
\X_{j_e}(\prodor{e^*\subset e} g_{e^*}^{\e(e,e^*)})
\end{equation}
where $g_{e^*}$ is understood to be $x_{e^*} h_{\t_{e^*}}
x_{e^*}^{-1}$ if $e^*\not\in\cT*$ and $x_{e^*} h_{\t^{\cT}_{e^*}}
x_{e^*}^{-1}$ if $e^*\in\cT^*$. This shows that this is the
evaluation of $L_\D$ where we colored the $e^*\not\in\cT^*$ by
$\O^*$, the $e^*\in\cT^*$ by $\t_{e^*}^{\cT^*}$, the $e\not\in\cT$
by $\O$ and the $e\in\cT$ by $j_e^{\cT}$. This is the first
result.

Now the second result can be proved using sliding identities. We
just proved that the LHS is the evaluation of a certain colored
link. We can apply sliding identities to this colored link. Our
claim is that the components corresponding to edges of $\cT$ and
dual edges of $\cT^*$ can be independently slid out of the link.
The consequence is that
$\le<\cO_{\D,\cT,\cT^*,j_e^\cT,\t_{e^*}^{\cT^*}}\ri>$ is equal to
the evaluation of a colored link made of the union of unknot
components carrying representations $j_e^\cT$ and
$\t^{\cT^*}_{e^*}$ and of the link $L_{\D,\cT,\cT^*}$ colored by
$\O$ and $\O^*$. The evaluation of this one gives $PR(\cM)$ while
the evaluation of the unknots components give the volume factor
$1$ and prove the formula.

To prove the sliding property, we will use the
specificity of the tree structure. We pick up an arbitrary vertex
that is called the root. Each vertex can be assigned a number
called its \textit{level}, and representing the number of edges
between it and the root (root has level 0). By extension, we
assign to an edge the level of its vertex of highest level.

We start with the tree $\cT^*$, and present the chain-mail link as
it arise from the triangulation, i.e to each tetrahedron is
associated a piece of link (see figure \ref{fig:chainmail}). Each
component of the link corresponding to faces (circles) carry a
color $\t$ if they are in $\cT$ or $\O^*$ if not. We will prove
that we can slide out the component of $\cT^*$ by proving the two
following statements
\begin{itemize}
\item A circle enclosing only edges of maximal level can be slid
out of the link; \item A circle enclosing edges of level $k$ can
be slid to enclose edges of level $k+1$;
\end{itemize}
Let us prove the first statement. If a circle enclose only edges
of maximal level, it means that around the target vertices of each
this edges, there is only $\O^*$ color - the contrary would mean
that there is another edge in the tree touching this vertices, and
contradicts the fact that we are at the maximum level. At each
such vertex we can slide the link this way
\begin{center}\psfrag{t}{$\t$}\psfrag{O}{$\O^*$}
\includegraphics[width=10cm]{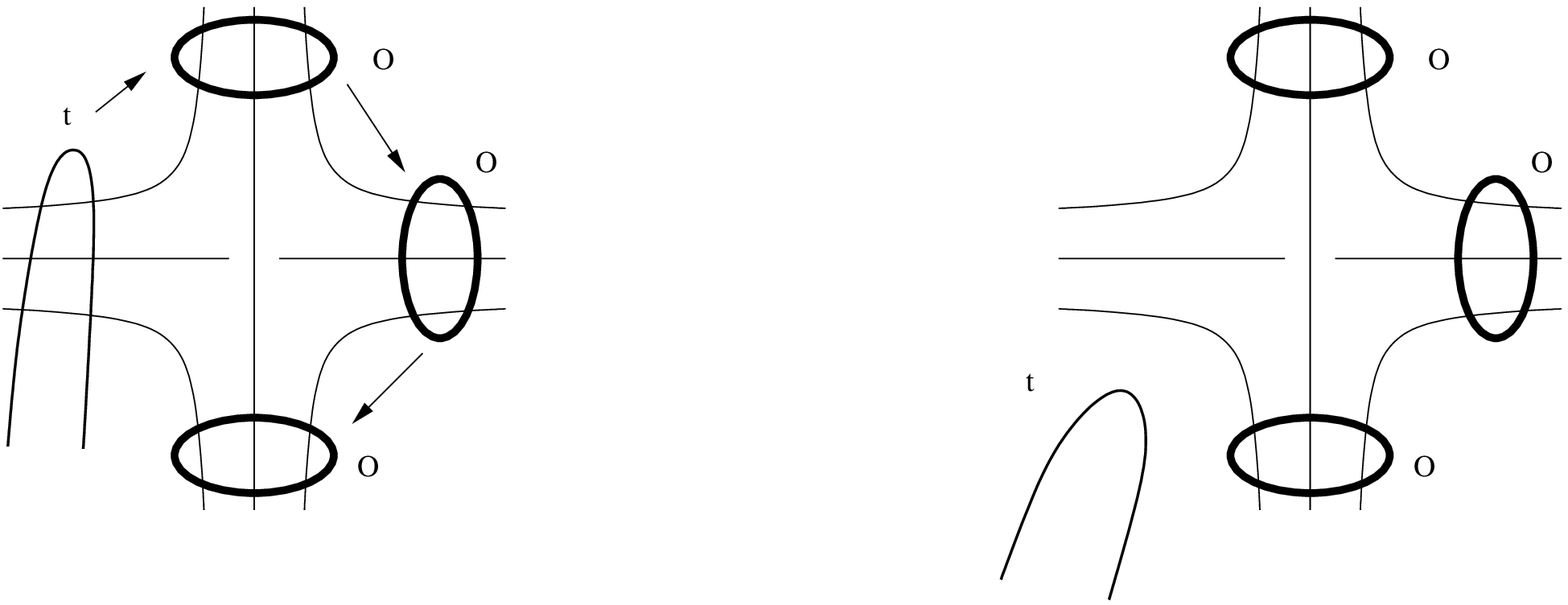}
\end{center}
The link can be thus liberated from all edges of maximal level.

Let us prove the second statement. If a link encloses an edge of
level $k$, around the target vertex there are edges which are not
in the tree, enclosed by circle with $\O^*$ representations, and
edges in the tree, which are by definition edges of level $k+1$.
One can thus slide our circle on the available $\O^*$
representations in order to make it enclose only edges of level
$k+1$ -- as in this example with only one $\O^*$ :
\begin{center}\psfrag{t}{$\t$}\psfrag{O}{$\O^*$}
\includegraphics[width=10cm]{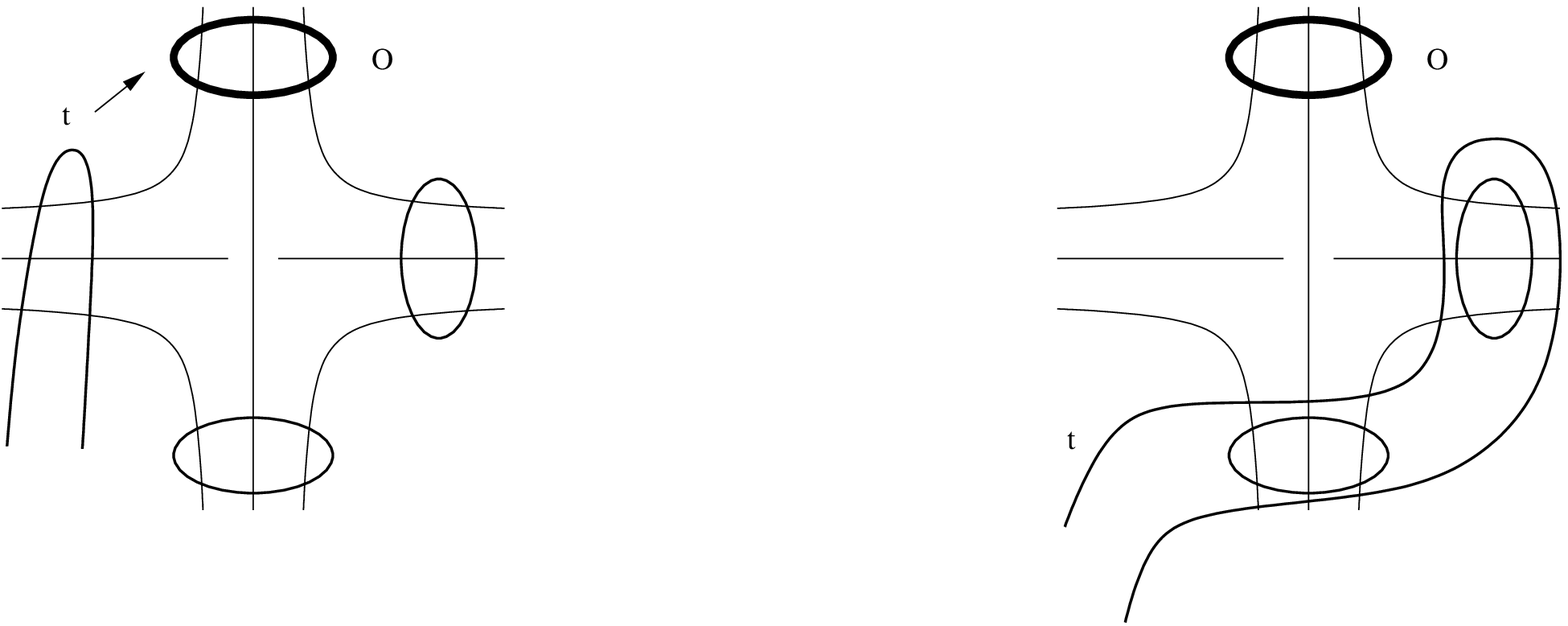}
\end{center}
The combination of the two statements show that every component of
the link corresponding to edges in $\cT^*$ can be slid out of
the link.

\begin{figure}[t]\psfrag{v}{$v$}
\includegraphics[width=4cm]{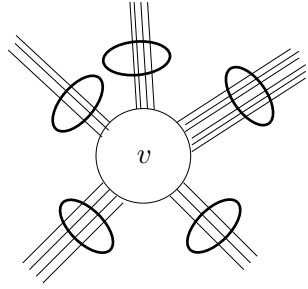} \caption{Piece of link
coming from a vertex of $\D$}
\label{fig:linkvertex}
\end{figure}

The same way, we can now present the link as a gluing of
elementary pieces associated to each vertex of $\D$ (see figure
\ref{fig:linkvertex}) and to same thing with the tree $T$. The
differences with the previous case being the fact that the vertex
can have any valence, and each edge is bounded by any number of
edges. However it does not affect the previous procedure. The key
point is that this procedure does not depend of whether some of
the component of the other type are still there or not, in other
words the procedure for $\cT^*$ and for $\cT$ do not interfere and
therefore the two operations commute.\ $\blacksquare$

\appendix

\section{Notations on $\SU(2)$}\label{app:su2}
We consider the group $\SU(2)$. We define the following $2 \times
2$ matrices
\begin{eqnarray}
h_\f&=&\mm{e^{i\f/2}}{0}{0}{e^{-i\f/2}} \\ a_\t&=&\mm{\cos \t/2
}{-\sin\t/2}{\sin\t/2}{\cos\t/2}
\end{eqnarray}
With these definitions, we consider the Euler-angle
parametrization of $\SU(2)$
\begin{equation}\label{eulerang}
g(\f,\t,\p)=h_\f a_\t h_\p,
\end{equation}
with the domain
\begin{eqnarray}
&& 0 \leq \f \leq 2\pi \\ && 0 \leq \t \leq \pi \\ && -2\pi \leq
\p \leq 2\pi.
\end{eqnarray}
The Cartan subgroup $H$ of $\SU(2)$ is $U(1)$ can be realized with
matrices $h_\p$ for $-2\pi \leq \p \leq 2\pi$ and the Weyl group
$W$ consists of the identity and the reflection $h_\phi
\rightarrow h_{-\phi}$. The quotient space $G/H$ can be realized
into $G$ using a choice of section. The Euler angle
parametrization gives the parametrization $x(\f,\t)=h_\f a_\t$ for
this section. The set $Conj$ of conjugacy class of $\SU(2)$ is
$[0,2\pi]$, and an explicit realization of it is also given by the
$h$ matrices. The representations of $\SU(2)$ are labelled by a
non-negative half-integer $j$ and realized on the carrier space
$V^j \sim \C^{2j+1}$. The representation is given by the Wigner
matrices $D^j(g)$. We consider the Haar measure $dg$ on $\SU(2)$.
The functions $\le\{\sqrt{d_j} D^j_{nn'}(g),\ 2j\in\N, -j \leq
n,n' \leq j \ri\}$ give an orthonormal basis of the space
$\cL^2(\SU(2),dg)$.
This orthonormality relation can be written
in terms of convolution product for characters
\be \int_G
\overline{\chi_{j'}(g)} \chi_j(gx) {dg} =
\d_{j,j'}\frac{ \chi_j(x)}{d_j}. \ee from which it is clear that
the group delta function can be written as
\begin{equation}
\d(G)=\sum_j {d_j} \X_j(G).
\end{equation}

Let us choose a  normalized Haar measure on the group, the
integration over the group can be written as an integration over
the conjugacy classes using the Weyl integration formula
\begin{equation}
\int_G dg f(g) = \int_{H/W} { \D(\t)^2} \left(\int_{G/H} f(xh_{\t}x^{-1})dx\right) \frac{d\t}{\pi}
\end{equation}
where $H=U(1)$ denote the Cartan subgroup,
 $\D(\t)=\sin(\t/2)$ and $W$ is the Weyl group, and
$dx$ is the normalized invariant measure on $G/H$.

We define the distribution $\delta_\phi(g)$ to be the distribution forcing
$g$ to be in the conjugacy class of $h_\phi$.
It is invariant under conjugation $\delta_\phi(g)= \delta_\phi(xgx^{-1})$
and normalized by
\be\label{dphi}
\int_G  \delta_\phi(g) f(g) dg =  \int_{G/H} f(xh_\phi x^{-1}) dx.
\ee
We can  write this distribution in terms of characters
\be
\delta_\phi(g)= \sum_j \chi_j(h_\phi)\chi_j(g).
\ee
The Weyl integration formula imply that
\be
\int_{H/W} \frac{d\phi}{\pi} \Delta^2(\phi) \delta_\phi(h_\theta) =1.
\ee
This means that we can relate this distribution to $\delta_\phi(\theta)$
the delta function on $H/W$,
\be\label{deltaHW}
  \delta_\phi(h_\theta)  =  \frac{\pi}{\Delta^2(\phi)}\delta_\phi(\theta).
\ee

\section{The double quantum group of $\SU(2)$} \label{app:dsu2}
\subsection{Drinfeld double of a group}
The Drinfeld double of a finite group $G$ is the quasi-triangular
algebra defined by the following settings
\begin{itemize}
\item Vector space structure : $\cD(G)=\cC(G)\otimes \C[G]$
with $\cC(G)$ the space of group functions and $\C(G)$ the group algebra.
A general element can be represented as a linear combination of
elements $(f\otimes g)$ where $f$ is a function over $G$ and $g\in
G$.
\item Product : $(f_1\otimes g_1)\cdot(f_2\otimes
g_2)=f_1(\cdot)f_2(g_1^{-1}\cdot g_1)\otimes g_1g_2$ \\ In
particular we have $(1\otimes g)\cdot(\d_k\otimes
e)=\d_k(g^{-1}\cdot g)\otimes g$ and $(\d_k\otimes
e)\cdot(1\otimes g)=\d_k\otimes g$.\item Coproduct : $\D(f\otimes
g)(x_1,x_2)=f(x_1x_2) g\otimes g$
\item Unit : $(1\otimes e)$;
\item Co-unit : $\e(f\otimes g)=f(e)$
\item Antipode : $S(f\ox
g)(x)=f(gx^{-1}g^{-1})\ox g^{-1}$
\item Star structure : $(f\otimes g)^* = \overline{f}(xgx^{-1})\otimes g^{-1}$
\item R-matrix : $R=\sum_g (\d_g\ox e)\ox(1\ox g)$, and in
particular $R^{-1}=\sum_g (\d_g\ox e)\ox(1\ox g^{-1})$
\end{itemize}
To give a ribbon structure to this Hopf algebra, we can consider
the Drinfeld element
\begin{equation}
u=\int dg S(1 \ox g)\cdot (\d_g \ox e) = \int dg (\d_g \ox
g^{-1}).
\end{equation}
which is central. This element satisfies $S(u)=u$ and $u$ itself
is thus a square root of $S(u)u$ that we chose as the ribbon
element.

In principle, for reasons of convergence, these definitions hold
only for the case of a finite group $G$. This construction has
been generalized to the case of locally compact group by
Bais, Koornwinder and Muller \cite{bais1}. For a group like
$\SU(2)$, all these settings can be translated using the
isomorphism $\cD(G)=\cC(G)\ox\C[G]\sim\cC(G\times G)$
\begin{eqnarray}
\cC(G)\ox\C[G]&\leftrightarrow& \cC(G\times G) \\ (f\ox g)
&\rightarrow& f(x)\d_g(y) \\ \sum_g F(\cdot,g)\ox g &\leftarrow&
F(x,y)
\end{eqnarray}
However, to keep the notations clearer, we will use the notations
with $\cD(G)=\cC(G)\ox\C[G]$, keeping in mind that all the
computations can be put under a rigorous form using this
isomorphism.
\subsection{Unitary Representations of  $\cD(\SU(2))$}
The classification of the representations of the double of a
compact group has been done in \cite{bais1, bais2}. The
representations of $\cD(\SU(2))$ are labelled by a pair made of a
conjugacy class and a representation of the corresponding
centralizer. In the case of $\SU(2)$, the conjugacy classes are
labelled by a $\t\in [0;2\pi]$, see appendix \ref{app:su2}. For
$\t\neq 0,2\pi$, the centralizer is $U(1)$, and the
representations $s$ of $U(1)$, $\rho_{s}(h_{\t})= e^{is\t}$
are such that $2s\in\mathbb{Z}$. If
$\t=0$, the centralizer is $\SU(2)$ itself. The representations of
$\cD(\SU(2))$ are thus of the form $(\t,s)$ or $(0,j)$. Among them
we will distinguish the so-called \textit{simple representations}.
They are representations of the form $(\t,0)$ or $(0,j)$.

\noindent\underline{Representations $(\t,s)$ :} \\%
The corresponding carrier space is given by
\begin{equation}
\stackrel{(\t,s)}{\cV}=\le\{\f\in\cL^2(G,\C) | \forall \xi \in
[-2\pi,2\pi], \f(xh_{\xi})=e^{is\xi}\f(x)\ri\},
\end{equation}
where the representation of an element $(f\otimes
g)\in\cD(\SU(2))$ is given by
\begin{equation}
\left[\stackrel{(\t,s)}{\Pi}(f\ox g)\f \right]( x) =
f(xh_\t x^{-1})\f(g^{-1}x).
\end{equation}
The ribbon element is acting diagonally
\begin{equation}
\stackrel{(\t,s)}{\Pi}(u) = e^{is\t}.
\end{equation}
An orthonormal basis of  $\stackrel{(\t,s)}{\cV}$ is
given by the Wigner functions (matrix elements of representations)
\bea\label{eqn:discretebasis}
\stackrel{(\t,s)}{\cV}=\bigoplus_{2(I-s)\in
\N}\stackrel{(\t,s)}{\cV}_I \\
\stackrel{(\t,s)}{\cV}_I= \{D^I_{ms}(x) ,\ -I\leq m \leq I\}
\eea
where the
matrix index $s$ is kept fixed.
The finite dimensional  subspaces $\stackrel{(\t,s)}{\cV}_I$ are invariant under
the $SU(2)$ subgroup of $\cD(\SU(2))$ composed of elements of the form
$1\otimes g$.

\vspace{1ex}
\noindent\underline{Representations $(0,j)$ :} \\ %
The carrier space of the representation $(0,j)$ is given by
\begin{equation}
\stackrel{(0,j)}{\cV}=\le\{\f\in\cC(G,V^j) | \forall g \in \SU(2),
\f(xg)=D^j(g^{-1})\f(x)\ri\}
\end{equation}
One can see that $\stackrel{(0,j)}{\cV}$ is actually isomorphic to
$V^j$ since such a function $\f$ is completely determined by its
value $\f(e)=v \in V^j$.
The ribbon element is acting trivially on in this representation.
An element $(f\otimes g)$ is represented
on $\stackrel{(0,j)}{\cV}\sim V^j$ by
\begin{equation}
\stackrel{(0,j)}{\Pi}(f\ox g) = f(e)D^j(g^{-1}).
\end{equation}


\section{Reshetikhin-Turaev evaluation over $\cD(\SU(2))$}\label{app:braidings}
In this appendix, we present the proofs of the main properties of
the chain-mail evaluation based on simple representations of
$\cD(\SU(2))$. In particular we present the computation of simple
colored tangles and of more involved tangles needed in our
constructions. We then expose the proofs for the basic properties
of chain-mail evaluation : the so-called sliding properties,
killing properties and fusion identities. Most of these results are new.

The representation spaces $\wmod{\t,s}$ are spaces of functions
over the group $SU(2)$. We will denote generically their elements
$\phi(x),\psi(y)...$ where $x,y,...\in\SU(2)$. We will thus write
the endomorphisms of these representation spaces by describing
their action on these generic elements. The representation spaces
$\wmod{0,j}$ are isomorphic to the representation spaces $V^j$ of
$SU(2)$. We will thus write the corresponding endomorphisms using
matrix elements of representations of $SU(2)$.
\subsection{Simples braidings}
We begin with the computation of simple braidings directly
obtained from the $R$-matrix of $\cD(\SU(2))$.
\begin{prop}[Simple braidings]
\begin{eqnarray}
{\psfrag{b1}{$\t$}\psfrag{b2}{$\a$}\figeq{\includegraphics[height=1cm]{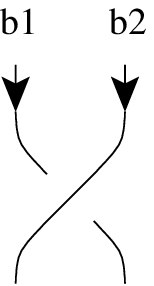}}}
&=& \phi(x)\otimes\psi(y)\to\psi(xh_\t^{-1} x^{-1}y)\otimes\phi(x) \\
{\psfrag{b1}{$\a$}\psfrag{b2}{$\t$}\figeq{\includegraphics[height=1cm]{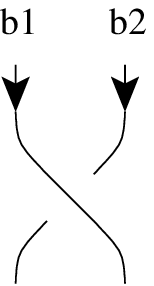}}}
 &=& \psi(y)\otimes\phi(x) \to \phi(x) \otimes \psi(xh_\t
x^{-1}y)\\
\figeq{\psfrag{b1}{$j_1$}\psfrag{b2}{$j_2$}\includegraphics[height=1cm]{braidelem.eps}}=
\figeq{\psfrag{b1}{$j_1$}\psfrag{b2}{$j_2$}\includegraphics[height=1cm]{braidelem.eps}}
&=& P_{12}
\end{eqnarray}
where $P_{12}$ denotes the permutation $V^{j_1}\ox V^{j_2} \to
V^{j_2}\ox V^{j_1}$ and the action is taken to go from top to bottom.
\end{prop}
\proof Let us prove the first braiding. The $R$-matrix of $\cD(\SU(2))$
is given by
\begin{equation}
R=\int_G dg\ (\d_g\ox e)\ox (1\ox g).
\end{equation}
its representation
on the space $\wmod{\t,0}\ox\wmod{\a,0}$ is, according
to the representations described in appendix \ref{app:dsu2}
\begin{equation}
\phi(x)\ox \psi(y) \to \int_G dg
\phi(x)\d_g(xh_\t x^{-1}) \psi(g^{-1}y) =
\phi(x)\otimes\psi(xh_\t^{-1}x^{-1}y)
\end{equation}
The braiding ${\cal R}= P R $ is obtained by composing with the permutation.
The second braiding is ${\cal R}^{-1}$ and
the other braidings are computed in the same way, by representing
the element $R$ on other representation spaces. \ \ $\blacksquare$
\subsection{Elementary tangles}
We now present the computation of an elementary tangle involving
trace over one representation.
\begin{prop}[Elementary tangles]
We have the following evaluations
\begin{equation}\label{elemtangle1}
 \figeq{\psfrag{t}{$j$}\psfrag{j}{$\t$}
\includegraphics[height=1.5cm]{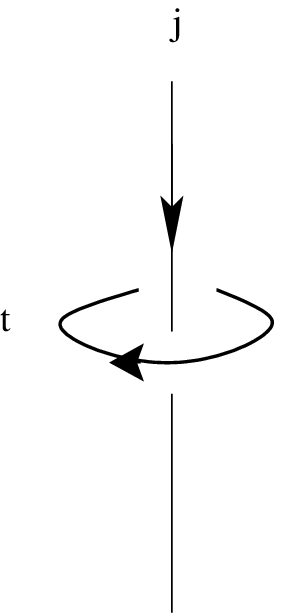}} = \phi(x) \to
\X_j(h_\t^{-1}) \phi(x) \ \ \ \ \
\figeq{\psfrag{t}{$j$}\psfrag{j}{$\t$}
\includegraphics[height=1.5cm]{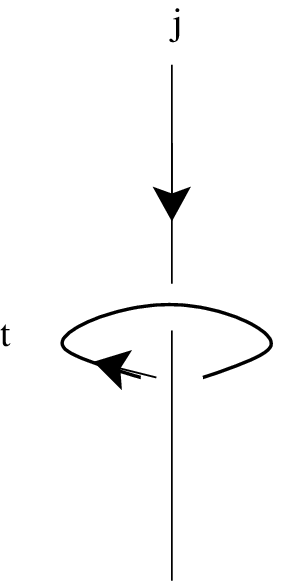}} = \phi(x) \to
\X_j(h_\t) \phi(x)
\end{equation}
\begin{equation}\label{elemtangle2}
\figeq{\psfrag{t}{$\t$}\psfrag{j}{$j$}
\includegraphics[height=1.5cm]{braid1.eps}} = \int dx
D^j(xh_\t^{-1} x^{-1}) \ \ \ \ \ \ \ \
\figeq{\psfrag{t}{$\t$}\psfrag{j}{$j$}
\includegraphics[height=1.5cm]{braid1inv.eps}} = \int dx
D^j(xh_\t x^{-1})
\end{equation}
\end{prop}
\proof Let us compute the evaluation of the following tangle
\begin{center}
\psfrag{j}{} \psfrag{t}{}
\includegraphics[width=4cm]{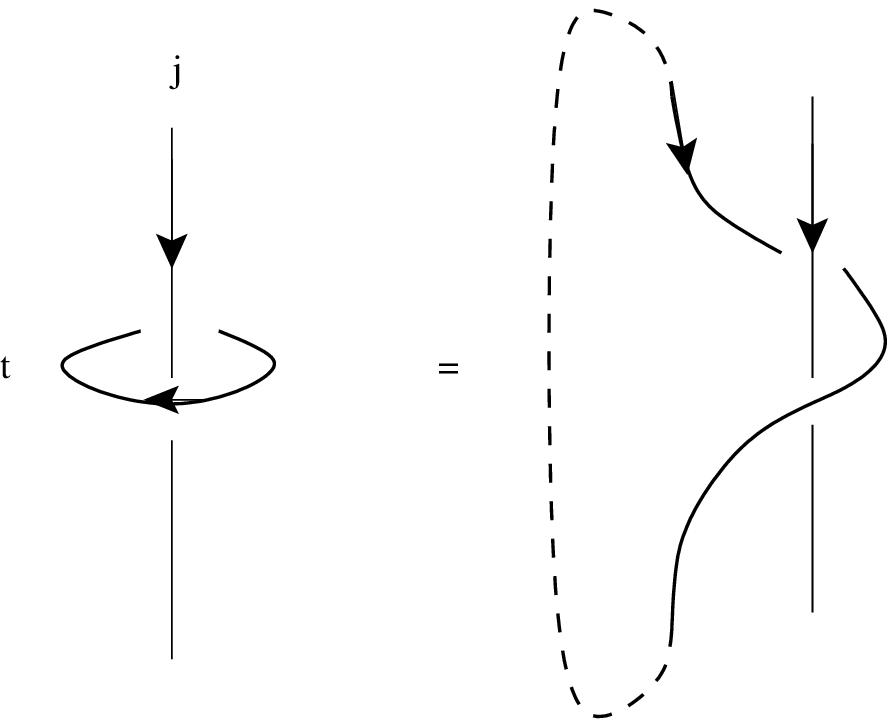}
\end{center}
We presented this tangle as the closure of a braid. The universal
element of $\cD(\SU(2))\ox\cD(\SU(2))$ associated to this braid is
\begin{eqnarray}\label{eqn:universalQ1}
Q ={\cal R}^2=R_{21}R_{12}&=&\int dg_1 dg_2 \le[(1\ox g_1)(\d_{g_2}\ox
e)\ri]\bigotimes \le[(\d_{g_1}\ox e)(1\ox g_2)\ri] \nonumber \\ %
&=&\int dg_1 dg_2 \le[\d_{g_2}(g_1^{-1}\cdot g_1)\ox
g_1\ri]\bigotimes \le[\d_{g_1}\ox g_2\ri] \label{eqn:absinv2strand}
\end{eqnarray}
where we use the rule of product of elements of $\cD(\SU(2))$, see
appendix \ref{app:dsu2}.
If we first represent $Q$ on the tensor product of
representation spaces $\wmod{0,j}\ox\wmod{\t,0}$. We get
\begin{eqnarray}
(\wrep{0,j}\ox\wrep{\t,0})\cdot Q :  v \otimes \phi(x) \to
D^{j}(xh_{\t}^{-1}x^{-1}) v \otimes \phi(x),
\end{eqnarray}
where $v$ is any vector in $ V^j$.
The trace over $V^j$ then gives the first result of (\ref{elemtangle1}.

We now represent $Q$ on the tensor product of representation
spaces $\wmod{\t,0}\ox\wmod{0,j}$.
\begin{eqnarray}
(\wrep{\t,0}\ox\wrep{0,j})\cdot Q^{(1)}   \f(x)\otimes v \to
\f(x) \otimes D^j(xh_\t^{-1}x^{-1})
\end{eqnarray}
To take the trace over the representation space $\wmod{\t,0}$, we
use its orthonormal basis given by the Wigner functions
$\le\{\sqrt{d_k}D^k_{m0}\ri\}$. We
want to prove that this endomorphism is diagonal and that the
reduced trace can be taken. Let us compute the trace over
$\wmod{\t,0}_k$ in the sense of def.\ref{def:redtrace}.
\begin{eqnarray}
\tr_{\wmod{\t,0}_k}\le[\wrep{\t,0}\ox\wrep{0,j}\cdot
Q\ri]&=&\sum_{m=-k}^k \int dx D^j(xh_\t^{-1} x^{-1}) d_k
D^k_{m0}(x)\overline{D^k_{m0}(x)}\\ &=& d_k \le(\int dx
D^j(xh_\t^{-1} x^{-1})\ri)
\end{eqnarray}
This shows that the endormorphism is diagonal and that its reduced
trace is $\int dx\ D^j(xh_{\t}^{-1}x^{-1})$. The other tangles are
evaluated in the same way.\ \ $\blacksquare$
\subsection{Complex colored tangles}
In this part, we present the evaluation of complex colored tangles
that are extensively used in our constructions.
\begin{prop}\label{prop:Testar}
\begin{equation} \label{Testar1}
\figeq{\psfrag{c}{$\t$}\psfrag{b1}{$j_1$}\psfrag{b2}{$j_2$}\psfrag{b3}{$j_3$}
\psfrag{bn}{$\!\!\!j_{N-1}$}\psfrag{bn1}{$\ \ j_N$}
\includegraphics[width=4cm]{braidn2.eps}}%
=\int_{G/H}dx\ \bigotimes_{i=1}^N D^{j_i}\le(xh^{-\e_i}_{\t} x^{-1}\ri)
\end{equation}
and
\begin{equation} \label{Testar2}
\figeq{\psfrag{c}{$j$}\psfrag{b1}{$\t_1$}\psfrag{b2}{$\t_2$}\psfrag{b3}{$\t_3$}\psfrag{bn}{ }\psfrag{bn1}{$\t_{n}$}
\includegraphics[width=3cm]{braidn2.eps}}%
=\otimes_{i=1}^n \phi_i(x_i) \to   \X_j\le(\overleftarrow{\prod}_{i=1}^n x_i
h_{\t_i}^{-\e_i}x_i^{-1} \ri) \otimes_{i=1}^n \phi_i(x_i)
\end{equation}
where $\e_i=\pm 1$ denotes the relative orientation of the $i$-th
strand and the circle component carrying representation $j$ (On
this picture the first strand is $\e=+1$ and the second $\e=-1$).
\end{prop}
\proof We denote $N$ the number of strands.
We represent the tangle as a closure of a braid, with the closure taking place on the first strand of the braid.
One can show that the action of the
universal element of $\cD(\SU(2))^{\ox(N+1)}$, representing the braid
associated to this tangle, on $\phi(x) \otimes_{i=1}^n \phi_i(y_i)$
is given by
\begin{eqnarray}\label{univQ}
\int_{SU(2)^{2N}} dg_1dg_2&& \prod_{i=1}^{N}dy_i  \delta_{g_1}(y_1\cdots y_n) \times \\
&&(\delta_{g_1g_2g_1^{-1}}\otimes g_1)\phi(x)\otimes
\bigotimes_{i=1}^N
\le\{
\begin{array}{c}
(\d_{y_i}\ox g_2)\ \mbox{ if }\ \e_i=+1 \\
(\delta_{g_2^{-1}y_i^{-1}g_2} \ox g_{2}^{-1})\ \mbox{ if }\ \e_i=-1
\end{array}\ri\}\phi_i(y_i)
\end{eqnarray}
The representation of this element on  the representation space
$\wmod{\t,s}\bigotimes_i \wmod{0,j_i}$ leads to
\begin{eqnarray}
\phi(x) \to &&\int_{SU(2)^{2N}} dg_1dg_2 \prod_{i=1}^{N}dy_i \delta_{g_1}(y_1\cdots y_n) \times\\
&&\phi(g_1^{-1}x) \delta_{g_1g_2g_1^{-1}}(xh_\t x^{-1})\otimes
\bigotimes_{i=1}^N
\le\{
\begin{array}{c}
\d_{y_i}(e) D^{j_i}( g_2^{-1})\ \mbox{ if }\ \e_i=+1 \\
\delta_{g_2^{-1}y_ig_2}(e) D^{j_i}(g_{2})\ \mbox{ if }\ \e_i=-1
\end{array}\ri\}.
\end{eqnarray}
The integrals can now be easily computed, giving
the value of the endomorphism of $\wmod{\t,s}\bigotimes_i \wmod{0,j_i}$
associated with the total braid.
\begin{equation}
\phi(x) \to \phi(x) \bigotimes_{i=1}^N D^{j_i}(xh_{\t}^{-\e_i}x^{-1})
\end{equation}
This endormorphism is diagonal on $\wmod{\t,0}$ and its reduced
trace is
\begin{equation}
\int_{G/H} dx\ \bigotimes_{i=1}^N D^{j_i}(xh_{\t}^{-\e_i}x^{-1}),
\end{equation}
which proves \ref{Testar1}.

If one now evaluate (\ref{univQ}) in the representation
$\wmod{0,j}\bigotimes_i \wmod{\theta_i,s_i}$ one gets
\begin{eqnarray}
\otimes_{i=1}^N \phi_i(x_i) \to &&
\int_{SU(2)^{2N}} dg_1dg_2 \prod_{i=1}^{N}dy_i \delta_{g_1}(y_1\cdots y_n)\times \\
&&D^j(g_1^{-1}) \delta_{g_1g_2g_1^{-1}(e)}
\otimes_{i=1}^N \delta_{y_i}(x_ih_{\t_i}^{\e_i}x_i^{-1}) \phi_i(x_i),
\end{eqnarray}
which gives after integration
\begin{eqnarray}
\otimes_{i=1}^N \phi(x_i) \to
D^j\le([\overrightarrow{\prod}_{i=1}^n x_i h_{\t_i}^{\e_i}x_i^{-1}]^{-1}\right)
\otimes_{i=1}^N \phi(x_i),
\end{eqnarray}
By taking the trace on $\wmod{0,j}$ we obtain the RHS of
(\ref{Testar2}).
\ \ $\blacksquare$
\vspace{1cm}

So far we considered only representations spaces of the form
$(0,j)$ or $(\t,0)$. Now we also consider the representation space
$(m,S)$, but for this one we will consider the discrete basis
(\ref{eqn:discretebasis}), whose elements are labelled by $(I,n)$
with $I-S\in\N$ and $-I\leq n \leq I$. With these notations, we
have the following evaluation
\begin{equation}
\figeq{\psfrag{t}{$j$}\psfrag{j}{$(m,S)$}
\includegraphics[height=2cm]{braid1.eps}}=\int_G du
\X^j(uh_m^{-1}u^{-1}) \sqrt{d_I}\sqrt{d_{I'}} D^I_{nS}(u) D^{I'}_{Sn'}(u^{-1}).
\end{equation}
This is proved by representing the element (\ref{eqn:universalQ1})
on $\wmod{0,j}\otimes\wmod{m,S}$. Let us compute the action of the
corresponding endomorphism on an element $\sqrt{d_I} D^I_{nS}$ of
$\wmod{m,S}$. We get
\begin{eqnarray}
&&\int dg_1 dg_2 \d_{g_2}(e) D^{j}(g_1^{-1})
\d_{g_1}(uh_mu^{-1})\sqrt{d_I} D^I_{nS}(g_2^{-1}u) \nonumber \\ %
&=& D^j(uh_m^{-1}u^{-1})\sqrt{d_I} D^I_{nS}(u).
\end{eqnarray}
We then take the trace over $\wmod{0,j}$ and get an endomorphism
of $(m,S)$. To express this endomorphism in the discrete basis of
$(m,S)$, we need to decompose the given result on this basis. We
get
\begin{equation}
\int_G du\ \X^j(uh_m^{-1}u^{-1})\sqrt{d_I d_{I'}} D^I_{nS}(u)
\overline{D^{I'}_{n'S}(u)}=\int_G du\ \X^j(uh_m^{-1}u^{-1})d_I
\sqrt{d_I d_{I'}}D^I_{nS}(u) D^{I'}_{Sn'}(u^{-1}).
\end{equation}
We can now state the following proposition, which is proved in the
same way :
\begin{prop}\label{prop:Testar2}
As before, we consider the representations spaces $\wmod{\t,0}$,
whose elements are functions $\phi(x)$ over $\SU(2)$ with the
covariance property $\phi(xh)=\phi(x),\ \forall h \in H$, and the
representation space $(m,S)$ with the discrete basis whose
elements are labelled by $(I,n)$. With these notations, we have
the following evaluation
\begin{equation}\label{Testar3}
\figeq{\psfrag{c}{$j$}\psfrag{b1}{$\t_1$}
\psfrag{b2}{$\t_2$}\psfrag{b3}{$\t_3$}\psfrag{bn}{$\t_{n}$}
\psfrag{bn1}{$(m,S)$}\includegraphics[width=3cm]{braidn2.eps}}%
=\prod_{i=1}^n \phi_i(x_i)\ \int_G du\ \X_j\le(\prod_{i=1}^n x_i
h_{\t_i}^{\e_i} x_i^{-1}\ uh_m u^{-1}\ri)
\sqrt{d_I}\sqrt{d_{I'}} D^I_{nS}(u) D^{I'}_{Sn'}(u^{-1})
\end{equation}
where $P_S$ denotes the projection over the basis vector $S$ of
$V^I$.
\end{prop}
\subsection{Killing properties}
The killing properties are
\begin{center}
\psfrag{Oj}{$\O$} \psfrag{Ot}{$\O^*$}
\psfrag{j}{$j$}\psfrag{t}{$\t$}
\psfrag{dj}{$\d_{j,0}$}\psfrag{dt}{$\d(h_\t)$}
\includegraphics[width=8cm]{killingids.eps}
\end{center}
Recall that $\O$ and $\O^*$ correspond to the formal linear
combinations of $(\t,0)$ and $(0,j)$ representations
\begin{eqnarray}
(\O^*) &=& \int_{H/W} \frac{d\t}{\pi} \D^2(\t) \ (\t,0) \\ %
(\O) &=& \sum_j d_j \ (0,j)
\end{eqnarray}
Using the evaluation (\ref{elemtangle2}) the LHS of the first identity can be written
\be
\int_{H/W} \frac{d\t}{\pi} \D^2(\t) \int_{G/H}dx D^j(xh_\t x^{-1})
=\int dg D^j(g) = \delta_{j0}
\ee
we where use that the integration is a  full group  integration by the
Weyl integration formula which projects over the trivial representation $j=0$,
we then get the first identity.
Using the evaluation
(\ref{elemtangle1}) and the Plancherel decomposition
\begin{equation}
\d(g)=\sum_j d_j \X^j(g),
\end{equation}
the LHS of the second graph is
\be
\delta(h_\t) Id_{\wmod{\t,0}},
\ee
and we get the second identity.

\subsection{Proof of the sliding properties}
The sliding properties are
\begin{center}
\psfrag{Oj}{$\O$} \psfrag{Ot}{$\O^*$}
\psfrag{j}{$j$}\psfrag{t}{$\t$}
\includegraphics[width=8cm]{slidingids.eps}
\end{center}
where \begin{eqnarray}
(\O)=\sum_j d_j\ (0,j), \,  (\O^*)= \int_H d\t \D(\t)^2\ (\t,0).
\end{eqnarray}
To prove them, we start from the braid
\begin{center}
\includegraphics[width=6cm]{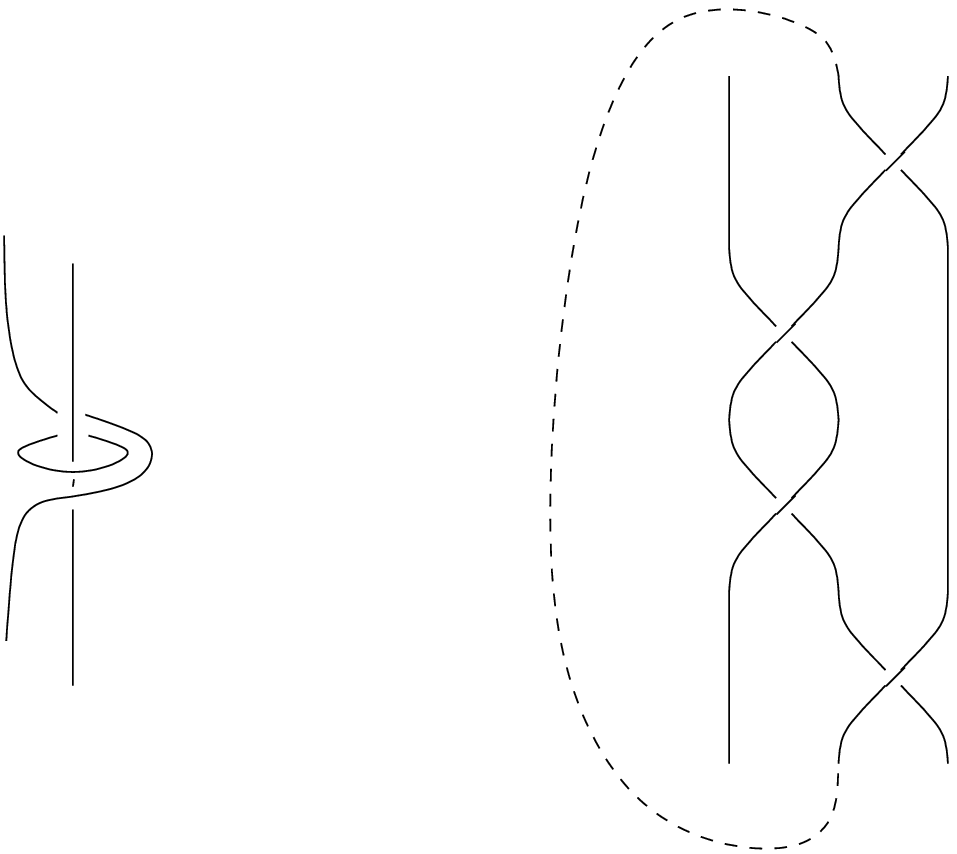}
\end{center}
The associated abstract element is
\begin{eqnarray}
P&=&\int \prod_{i=1}^4\ %
\le[(1\ox g_2)(\d_{g_3}\ox e)\ri]\ox%
\le[(1\ox g_1)(\d_{g_4}\ox e)\ri]\ox%
\le[(\d_{g_1}\ox e)(\d_{g_2}\ox e)(1\ox g_3)(1\ox g_4)\ri]
\nonumber \\ &=& \int dg_2 dg_3 dg_4 \le[\d_{g_2g_3g_2^{-1}}\ox g_2\ri]
\ox \le[\d_{g_2g_4g_2^{-1}}\ox
g_2\ri] \ox \le[\d_{g_2}\ox g_3g_4\ri]
\end{eqnarray}
where use the product rules in $\cD(\SU(2))$.
We first represent this element on
$\wmod{0,j}\ox\wmod{0,k}\ox\wmod{\t,s}$
\begin{equation}
\le[(\wrep{0,j}\ox\wrep{0,k}\ox\wrep{\t,s})\cdot P \ri]\f(x)=%
D^j(xh_\t^{-1} x^{-1})\otimes D^k(xh_\t^{-1}x^{-1})\f( x)
\end{equation}
We  take the trace over  $k$ representations
and sum over them with the factor $d_k$. We get
\begin{equation}
\d(h_\t) D^j(1) \ox Id_{\wmod{\t,s}}
\end{equation}
which is the LHS of the first sliding identity.
To prove the second identity, we represent $P$ over
$\wmod{\t_1,0}\ox\wmod{\t_2,0}\ox\wmod{0,j}$, we obtain
\begin{equation}
\phi_1(x_1)\ox \phi_2(x_2)
\to
\phi_1(x_1)\ox \phi_2(x_2)\ox D^j(x_1h_{\t_1}^{-1}x_1^{-1})D^j(x_2h_{\t_2}^{-1}x_2^{-1})
\end{equation}
Tracing over $\wmod{\t_2,0}$ (according to the reduce trace
prescription) gives the endomorphism
\be
\phi(x)\to \chi_j(h_{\t_2}) \phi(x) \ox D^j( xh_{\t_1}x^{-1}).
\ee
Integrating over the $\t_2$ allows
to built $g_2=x_2h_{\t_2}x_2^{-1}$
\begin{equation}
\int dg_2 D^j(x_1h_{\t_1}^{-1}x_1^{-1})\d_{y_1}(x_1)D^j(g_2) =
\d_{y_1}(x_1)
\end{equation}
and prove the second sliding identity. It should be noted that for
this identity, the Lickorish proof \cite{lickorish} does not work
since the fusion of two $(\t,0)$ representations contains also
non-simple representations. \ $\blacksquare$

\subsection{Fusion properties and Clebsh-Gordan coefficient}
In this section we recall the construction  \cite{bais1,bais2} of Clebsch-Gordan
coefficient associated with simple representations of $D(SU(2))$ and then establish the
 orthonormality property satisfied by these coefficients.

\subsubsection{Tensor product $(0,j_1)\otimes (0,j_2)$}
The tensor product decomposition of simple representations of type
$(0,j)$ is given
\begin{equation}
(0,j_1)\otimes(0,j_2) = \bigoplus_{j_3=|j_1-j_2|}^{j_1+j_2}
(0,j_3).
\end{equation}
Each representation on the RHS appears with multiplicity 1, and
the corresponding Clebsh-Gordan coefficient is given by the
$SU(2)$ normalized Clebsh-Gordan coefficient
\begin{equation}
\cC^{(0,j_1)(0,j_2)(0,j_3)}_{m_1m_2m_3}=C^{j_1j_2j_3}_{m_1m_2m_3}.
\end{equation}
Let us remark that simple representations of type $(0,j)$
decompose only on simple representations of the same type.
\subsubsection{Tensor product $(\t_1,0)\otimes (\t_2,0)$}
The tensor product of two simple representations of type $(\t,0)$
is more involved. The tensor product $(\t_1,0)\otimes(\t_2,0)$
decomposes on representations $(\t_3,n_3)$ with multiplicity 1.

We want to express the corresponding Clebsh-Gordan coefficient
\begin{equation}
\psfrag{r1}{$(\t_1,0)$}\psfrag{r2}{$(\t_2,0)$}\psfrag{r3}{$(\t_3,n_3)$}
\psfrag{x1}{$x_1$}\psfrag{x2}{$x_2$}\psfrag{x3}{$x_3$}
\figeq{\includegraphics[height=2cm]{clebsh.eps}}=\cC^{(\t_1,0)(\t_2,0)(\t_3,n_3)}_{x_1x_2x_3}.
\end{equation}
This coefficient defines an invariant map $\wmod{\t_1,0}\ox \wmod{\t_2,0}\to\wmod{\t_3,s_3}$
\be
\phi_1(x_1)\ox \phi_2(x_2) \to \int dx_1 dx_2\cC^{(\t_1,0)(\t_2,0)(\t_3,n_3)}_{x_1x_2x_3}
\phi_1(x_1) \phi_2(x_2).
\ee
 described as follows. Consider three
conjugacy class $\t_1,\t_2,\t_3$, satisfying
\be\label{rel}
|\t_1-\t_2|\leq \t_3 \leq \t_1+\t_2, \, \t_1+\t_2+\t_3 \leq 4\pi.
\ee
It exists a unique $\t\in U(1)\backslash \SU(2)/U(1)\sim [0,\pi]$
such that $\t_3$ is the
conjugacy class of $h_{\t_1}a_\t h_{\t_2}a_\t^{-1}$ (see appendix
\ref{app:su2} for the notation $a_\t$). This $\t$ is a function of
$\t_1,\t_2,\t_3$ and parametrizes the relation between the 3
conjugacy classes. Taking the trace of this defining relation we
get
\begin{equation}\label{eqn:tt3}
\cos\frac{\t_3}{2}=\cos\frac{\t_1}{2}\cos\frac{\t_2}{2}-\cos\t\sin\frac{\t_1}{2}\sin\frac{\t_2}{2}.
\end{equation}
Now let us denote by $w_\t$ the unique (up to a choice of
section) element of $G/H$ such that
\begin{equation}
h_{\t_1}a_\t h_{\t_2}a_\t^{-1}=w_\t h_{\t_3}w_\t^{-1}
\end{equation}
From \cite{bais1} we know  that the Clebsh-Gordan
coefficient is given by
\begin{eqnarray}
\cC^{(\t_1,s_1)(\t_2,s_2)(\t_3,s_3)}_{x_1x_2x_3}&= &
(4\sin(\t_1/2)\sin(\t_2/2)\sin(\t_3/2))^{-1/2}\times \\
&& \int_{U(1)} d\xi_1d\xi_2d\xi_3\,
e^{-is_3\xi_3} \d_{x_3h_{\xi_3}}(x_1 h_{\xi_1} w_\t)\ \d_{x_2h_{\xi_2}}(x_1h_{\xi_1}a_\t)
\end{eqnarray}
where $d\xi$ is normalized on $U(1)$ and the normalization coefficient is chosen in order to simplify the
orthogonality relations. Let us recall that in this
expression, $\t,a_\t$ and $w_\t$ are uniquely determined in terms of
$\t_1,\t_2$ and $\t_3$.
If $\t_1,\t_2,\t_3$ do not satisfy the relation (\ref{rel}) we define the Clebsh-Gordan coefficient to
be zero.
\begin{prop}\label{prop:orthoclebsh}
We have the orthogonality relation
\begin{equation}
\int_{0}^{2\pi} \sin(\t_3/2)^2 d\t_3 \sum_{n_3\in\Z} \int_{G/H} dx_3\
\cC^{(\t_1,0)(\t_2,0)(\t_3,n_3)}_{x_1x_2x_3}
\overline{\cC^{(\t_1,0)(\t_2,0)(\t_3,n_3)}_{z_1z_2x_3}}=\bar\d_{z_1}(x_1)\bar\d_{z_2}(x_2),
\end{equation}
with
\begin{equation}\label{eqn:deltabasis}
\bar{\d}_x(g) \equiv \int_H dh \d_{xh}(g).
\end{equation}
\end{prop}
\proof The LHS is
\begin{eqnarray}
 \int_{0}^{2\pi} d\nu(\t_3) \int_{G/H} dx_3 \sum_{n_3}&& \int_H \prod_{i=1}^3 d\xi_i \
e^{in_3\xi_3}
\d_{x_1h_{\xi_1}}(x_3h_{\xi_3}w_\t^{-1}) \d_{x_2h_{\xi_2}}(x_1h_{\xi_1}a_\t) \nonumber
\\ && \int_H \prod_{i=1}^3 d\xi'_i \ e^{-in_3\xi'_3}
\d_{z_1h_{\xi'_1}}(x_3h_{\xi'_3}w_\t^{-1}) \d_{z_2h_{\xi'_2}}(z_1h_{\xi'_1}a_\t),
\end{eqnarray}
with
\be\label{meas}
d\nu(\t_3)=\frac{\sin(\t_3/2)}{ 4  \sin(\t_1/2)\sin(\t_2/2)} d\t_3
\ee
We perform the sum over $n_3$, this imposes $\d(\xi_3-\xi'_3)$, we then
rearrange the arguments of the delta functions and the
result can be written as
\begin{equation}
\int_{0}^{2\pi}  d\nu(\t_3)
\int_{G/H} dx_3  \int_H d\xi_1 d\xi_2 d\xi_3
\d_{x_1h_{\xi_1}}(x_3h_{\xi_3}w_\t^{-1})\d_{x_2h_{xi_2}}(x_1h_{\xi_1}a_\t)
\bar\d_{z_1}(x_1)\bar\d_{z_2}(x_2)
\end{equation}
We can put together the variables $x_3\in G/H$ and $\xi_3\in H$ to
build an element of $G$ and integrate the first delta function. It remains
\begin{equation}\label{deltaid}
\int_{0}^{2\pi} d\nu(\t_3)
\int_H d\xi_1d\xi_2 \d_{x_1^{-1}x_2}(h_{\xi_1}a_\t h_{\xi_2}^{-1})
\bar\d_{z_1}(x_1)\bar\d_{z_2}(x_2)
\end{equation}
The delta function $\d_{x_1^{-1}x_2}(h_{\xi_1}a(\t)h_{\xi_2}^{-1})$
imposes $\t_3$ to be the conjucacy class of
$x_1h_{\t_1}x_1^{-1}x_2h_{\t_2}x_2^{-1}$.
In order to compute the integral we need  to
compute the jacobian of the transformation from $\t$ to $\t_3$. This
jacobian from $\t$ to $\t_3$ is easily computed from
the relations  (\ref{eqn:tt3}, \ref{meas}) and reads
\begin{equation}
{d\nu(\t_3)}= \frac{\sin\t}{2} {d\t}.
\end{equation}
The integral in equation (\ref{deltaid})
reads
\be
\int_0^\pi \frac{sin\t}{2}d\t \int_H d\xi_1 d\xi_2 \d_{x_1^{-1}x_2}(h_{\xi_1}a(\t)h_{\xi_2}^{-1})
= \int_G dg \d_{x_1^{-1}x_2}(g)=1
\ee
where we have recognized the normalized Haar measure written in terms of the Euler angles
(\ref{eulerang}).
\ $\blacksquare$
\subsection{Fusion properties}
We prove here the two fusion identities for simple representations
of the same kind.
\subsubsection{Fusion properties of the $(0,j)$ representations}
The fusion identity for $(0,j_1)\ox (0,j_2)$ reads
\begin{equation}\label{fus1}
\psfrag{t}{$\t$}\psfrag{j1}{$j_1$}\psfrag{j2}{$j_2$}
\psfrag{r1}{$j_1$} \psfrag{r2}{$j_2$} \psfrag{r3}{$j_3$}
\psfrag{R}{$\t$}
\figeq{\includegraphics[height=3cm]{braid2.eps}}=\sum_{j_3}d_{j_3}
\figeq{\includegraphics[height=3cm]{fusion21.eps}}
\end{equation}
To prove it, we consider the following evaluation of the LHS
\begin{equation}
\int dx \le[D^{j_1}\ox D^{j_2}\ri](xh_\t^{-1}x^{-1})
\end{equation}
We can decompose the $\SU(2)$ representations into Clebsh-Gordan
series
\begin{equation}\label{fusproof}
\sum_{j3} d_{j_3}\int dx\
\le[C^{j_1j_2j_3}D^{j_3}(xh_\t^{-1}x^{-1}) \overline{C^{j_1j_2j_3}}\ri]
\end{equation}
where $C^{j_1j_2j_3}$ are $\SU(2)$ normalized Clebsh-Gordan coefficients.
Due
to the fact that the fusion of two representations $(0,j_1)$ and
$(0,j_2)$ decomposes only on the $(0,j)$ representations, with
Clebsh-Gordan coefficients $\cC^{j_1j_2j_3}= C^{j_1j_2j_3}$ which are those of
$\SU(2)$, (\ref{fusproof}) is the RHS of (\ref{fus1}).
This result can be extended to the fusion of an arbitrary number of
$(0,j)$.

\subsubsection{Fusion of two $\t$}
The fusion identity for $(\t_1,0)\ox(\t_2,0)$ reads
\begin{equation}
\psfrag{t}{$j$}\psfrag{j1}{$\t_1$}\psfrag{j2}{$\t_2$}
\figeq{\includegraphics[height=3cm]{braid2.eps}} = \int d\v(\t_3)
\sum_{s_3} \psfrag{r1}{$(\t_1,0)$}\psfrag{r2}{$(\t_2,0)$}
\psfrag{r3}{$(\t_3,s_3)$}\psfrag{R}{$(0,j)$}
\figeq{\includegraphics[width=2cm]{fusion21.eps}}
\end{equation}
where
\begin{equation}
\psfrag{r1}{$\t_1$}\psfrag{r2}{$\t_2$}\psfrag{r3}{$(\t_3,s_3)$}
\psfrag{x1}{$x_1$}\psfrag{x2}{$x_2$}\psfrag{x3}{$x_3$}
\figeq{\includegraphics[width=2cm]{clebsh.eps}}=\cC_{x_1\ \ x_2 \
\ x_3 \ \ }^{(\t_1,0)(\t_2,0)(\t_3,s_3)}
\end{equation}
denotes the Clebsh-Gordan coefficient of the tensor product of
representations $(\t_1,0)\ox(\t_2,0)$ into $(\t_3,s_3)$
\footnote{Note that contrary to $(0,j)$ case, the tensor product
of two $(\t,0)$ representations does not decompose only on simple
representations.}, and $\v(\t_3)$ the measure (explicit expression
for the coefficient and the measure are given in appendix
\ref{app:dsu2}).
We have already computed the LHS in proposition{prop:Testar1}, its action on
$\phi_1(x_1)\otimes \phi_2(x_2)$ gives
\begin{equation}
\X_j(x_1h_{\t_1}^{-1}x_1^{-1}x_2h_{\t_2}^{-1}x_2^{-1})
\f_1(x_1)\otimes \f_2(x_2)
\end{equation}
 On the basis of delta functions, the evaluation of
the RHS is then
\begin{equation}
\X_j(x_1h_{\t_1}^{-1}x_1^{-1}x_2h_{\t_2}^{-1}x_2^{-1})\bar\d_{y_1}(x_1)\bar\d_{y_2}(x_2)
\end{equation}
Now looking at the proof of orthogonality of the Clebsh-Gordan
coefficients (see Prop. \ref{prop:orthoclebsh}), we can easily
modify this proof, inserting a $\X_j(\t_3)$ functions, to get the
following evaluation of the LHS
\begin{eqnarray}
\int d\t_3 \Delta(\t_3) \sum_{n_3} \int_{G/H} dx_3\
 & \cC^{(\t_1,0)(\t_2,0)(\t_3,n_3)}_{x_1x_2x_3}\ \X_j(\t_3)\
\overline{\cC^{(\t_1,0)(\t_2,0)(\t_3,n_3)}_{z_1z_2x_3}}= \\
&\X_j(x_1h_{\t_1}x_1^{-1}x_2h_{\t_2}x_2^{-1})\bar\d_{z_1}(x_1)\bar\d_{z_2}(x_2)
\end{eqnarray}
which  proves the identity.
\ $\blacksquare$

\section{Proof of Lemma 1}\label{lemma1}

Note that  $\cT^{*}$ possess  $ |\D_{3}|+|\bD_{2}|$ vertices and
$ |\D_{3}|+|\bD_{2}|-1$ edges.
Also $|\tilde{\D}_{1}^{*}|= |{\D}_{1}^{*}|+|\bar{\D}_{1}^{*}| =
|{\D}_{2}|+|\bar{\D}_{1}|$.
Therefore the  cardinal of $\wD_{\cT^{*}}= \wD^*_1 \backslash
\cT^{*}$ is given by
\be\label{cardD*}
|\wD_{\cT^{*}}|= |\D_{2}|+ |\bD_{1}|-|\D_{3}|-|\bD_{2}|+1.
\ee

$\cT$ possesses   $ |\D_{0}|-|\bD_{0}|+1$ vertices and
$ |\D_{0}|-|\bD_{0}|$ edges if
$\partial M= \coprod_{i=1}^n \Sigma_i$ is non empty.
The cardinal of $\D_{\cT}$ is then
given by\footnote{If $\partial M =\empty$, $\cT$ possesses one edge
less
and $|\D_{\cT}|= |\D_{1}|-|\D_{0}|+|\bD_{0}| +1$}
\be\label{cardD}
|\D_{\cT}|= |\D_{1}|-|\D_{0}|+|\bD_{0}|
\ee

the difference LHS of (\ref{diffg})  gives
\be\label{diff0}
|\wD_{\cT^{*}}|- |\D_{\cT}|= \X(M) -\X(\partial M) +1 ,
\ee
where $\X(\Sigma)$ denotes the Euler characteristic
\be
\X(M)=\sum_{i=0}^3 (-1)^i|\D_i|, \, \X(\partial M)=\sum_{i=0}^2
(-1)^i|\bD_i|.
\ee
It is well known that for an open 3 dimensional manifold we have
\be
2\X(M)- \X(\partial M)= \X(M\sharp_{\partial M} \bar{M})=0.
\ee
Moreover,
\be
\X(\partial M)= \X(\coprod_i\Sigma_i)=   \sum_i \X(\Sigma_i).
=\sum_i(2-2g_{\Sigma_i}).
\ee
All this result imply that (\ref{diff0}) is equal to
$1+ \sum_i (g_{\Sigma_i}-1)$ which proves the second part of the
lemma.

In order to prove the first part of the Lemma
we need to use the {\it Bianchi Identity}, whose proof is recalled
in lemma \ref{Bianchid} of appendix \ref{app:proofinv}:
For every set of dual faces $f^*$ forming a closed surface $\cS^*$
having the topology of a 2-sphere, it exists some group elements
$k_{f^*}$ (functions of the $g_{e^*}$), some $\s_{f^*}=\pm 1$ and
an order of the faces $f^*$ such that
\begin{equation}
\prodor{f^*\in\cS^*} k_{f^*} G_{f^*}^{\s_{f^*}} k_{f^*}^{-1}=1
\end{equation}
We now chose  the connection to satisfy $G_{\D_{d}} =1$.
Suppose that $f^*\in \bD^*_2$ is a boundary face.
This face belongs to a unique elementary three cell $c^*\in \wD_3^*$.
If this three cell is not intersecting the tree $T$
then all group elements, except $G_{f^*}$, associated to faces
belonging to
$c^*$ are equal to one if $G_{\D_{\cT,\cT^*}} =1$.
By the Bianchi identity this means that $G_{f^*}=1$.

If $c^*$ intersects the tree, $f^*$ is said to be connex to the
tree\footnote{By construction there is only one
face with such property}.
In this case, one can consider the surface
$\sigma_T \in \wD_2^*$ which is the boundary
of the tubular neighborhood of $T$.
Explicitly  this surface can be built as the union of boundary faces
connex to the tree
and the union of the $f^*$ for $f^*\sim e$, with $e$ touching the
vertices of $T$ without
belonging to $T$.
Since  $T$ is a tree of $\D$, its tubular neighborhood has the
topology of a 3-ball\footnote{This is true only if $\D$ is a
triangulation of a real manifold (not a
pseudo-manifold). In this case the neighborhood of every vertex $v$
is a ball} and
its boundary $\sigma_T$ has the topology of a 2-sphere.
We can then use the Bianchi identity on $\Sigma_T$ to conclude
that $G_{f^*}=1$.

Now that we have proven $G_{f^*}=1$  for all boundary faces
we consider $f^*\in \D^*_2$  an internal dual face
intersecting $T$ ($f^*\sim e$, $e\in T$).
$f^*$ cut the tree tubular neighborhood surface $\sigma_T$ in two
halves
$\Sigma_{T,f^*},\widetilde{\Sigma}_{T,f^*}$.
Moreover by the previous argument and the hypothesis $G_{\D_{d}} =1$
all group elements except $G_{f^*}$ are equal to one.
Using the Bianchi identity on $\Sigma_{T,f^*}$ one conclude
$G_{f^*}=1$.
This conclude the proof of the Lemma.


\section{Proof of the invariance of the Ponzano-Regge
invariant}\label{app:proofinv}

In this appendix, we prove that the Ponzano-Regge invariant
defined by equation (\ref{eqn:defregulPR}) is independent of all
the ingredients and thus is actually an invariant $PR(\cM)$ of the
manifold $\cM$.

\subsection{Independence from the orientations}

\begin{prop}\label{prop:invorient}
The definition of $PR[\cM,\D,\cT,\cT^*]$ does not depend on the
choice of the orientations of edges, faces, and of the choice of
starting dual vertex $st(e)$.
\end{prop}
\proof{
Let us reverse the orientation of an edge $e$ of $\D$. This will
change the signs of all the $\e(e,e^*)$ for the $e^*\subset e$ and
the order of the product. The effect on the integrand is to
reverse $G_e$ into $G_e^{-1}$. This does not affect the value of
the integral since $\d(G^{-1})=\d(G)$ for every group element $G$.

Let us reverse the orientation of a dual edge $e^*$. This will
change the signs of all the $\e(e^*,e)$ for $e\supset e^*$. The
effect on the integrand is to reverse every $g_{e^*}$ into
$g_{e^*}^{-1}$. This does not affect the integral since the Haar
measure is invariant by inversion $dg^{-1}=dg$.

Let us choose another starting vertex $st(f^*)$ for a given dual
face $f^*\sim e$. The effect is to conjugate $G_e$ by adjoint
action of a group element. This does not affect the integral since
$\d(k^{-1}G_ek)=\d(G_e)$.}

\subsection{Independence from choice of trees}

To prove the invariance under changes of trees, we will need the
following lemma regarding the relation between different choices
of maximal trees \cite{FLnoncomp}.
\begin{lemma}[Elementary moves of maximal
tree]\label{lemma:movetrees} Two homologous maximal trees $T_1$ and $T_2$ in
a graph $\G$  be related by a sequence of elementary
moves. $T_2$ is said to be related to $T_1$ by an elementary move
if $T_2$ is obtained from $T_1$ in the following way : Pick up a
vertex $v$ in $T_1$, an edge $\e$ in $T_1$ and an edge $\te$ not
in $T_1$, both edges touching $v$. Create $T_2$ by removing $\e$
from $T_1$ and add $\te$ to it.
\end{lemma}

\vspace{1ex}

\begin{prop}\label{prop:invtreestar}
The definition of $PR[\cM,\D,\cT,\cT^*]$ does not depend on the
choice of $\cT^*$.
\end{prop}
\proof {The proof of the independence in the choice of $\cT^*$ is
based on two ingredients. The first one is the invariance of the
Haar measure by left (right) translation and inverse. The second
one is the particular structure of the partition function i.e the
way the $G_e$ are built from the $g_{e^*}$.

We consider first the invariance under an elementary
move of the tree $\cT^*$. We pick up a vertex $v^*$, an edge
$\e^*\in\cT$ and an edge $\te^*\not\in\cT^*$, see figure
\ref{fig:movetreeTstar}. We want to perform the move that remove
$\e^*$ from $\cT^*$ and add $\te^*$ to it.
\begin{figure}[t]\psfrag{R}{$R^*$}\psfrag{ep}{$\e^*$}\psfrag{te}{$\te^*$}\psfrag{w}{}\psfrag{v}{$v^*$}

\includegraphics[width=4cm,height=6cm]{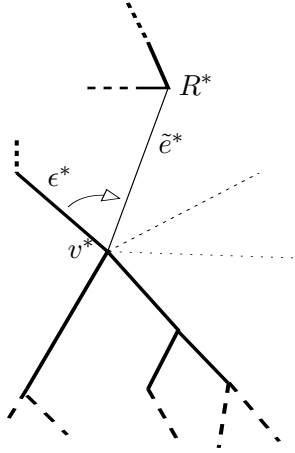}
\caption{Elementary move of $\cT^*$ at a vertex $v^*$. The tree is
in thick lines. }\label{fig:movetreeTstar}
\end{figure}
In terms of the explicit expression, we want to prove that
\begin{eqnarray}
&&\int_G \prod_{e^*} dg_{e^*}\  \le(\prod_{e^*\in\cT^*,\
e^*\neq\e^*} \d(g_{e^*}) \ri)\ \d(g_{\e^*})\  \prod_{e\not\in\cT}
\d(G_e)\nonumber \\ %
&=&\int_G \prod_{e^*} dg_{e^*}\ \le(\prod_{e^*\in\cT^*,\
e^*\neq\e^*} \d(g_{e^*}) \ri)\ \d(g_{\te^*})\ \prod_{e\not\in\cT}
\d(G_e) \label{eqn:moveTstar}
\end{eqnarray}
The strategy is to start from the first expression, to perform a
change of variable and to show that it leads to the second
expression.

\underline{Change of variable :} For simplicity, we will consider
that all the edges touching $v^*$ are oriented toward $v^*$. We
denote $R^*$ the other vertex of $\te^*$. We consider it as the
root of the tree i.e. it induces a partial order on the vertices,
in particular we will talk about the vertices $w^*\preceq v^*$
which are \textit{under} $v^*$ in the tree (including $v^*$
itself). We define $\s(e^*)=1$ if the source of $e^*$ is a vertex
$w^*\preceq v^*$, and $\s(e^*)=0$ if not. We define $\tau(e^*)=1$
if the target of $e^*$ is a vertex $w^*\preceq v^*$, and
$\tau(e^*)=0$ if not. We perform the following change of variable
\begin{eqnarray}
g_{\e^*}&\to& g_{\e^*} \nonumber \\ %
g_{\te^*}&\to& g_{\e^*} g_{\te^*}^{-1} g_{\e^*} \nonumber \\ %
\forall e^* \mbox{\ s.t.\ } e^*\neq \e^*, e^*\neq \te^*,\ \
g_{e^*} &\to& k^{-\s(e^*)} g_{e^*}\ k^{\tau(e^*)},\ \ \
\mbox{with}\ \ k\equiv g_{\te^*}^{-1}g_{\e^*}
\label{eqn:changevariable}
\end{eqnarray}

\underline{Effect of the change of variable :} Let us examine the
effect of the change of variable on the integral expression. Our
claim is the following
\begin{eqnarray}
&&\int_G \prod_{e^*} dg_{e^*}\
\le(\prod_{\stackindex{e^*\in\cT^*}{e^*\neq\e^*}} \d(g_{e^*})
\ri)\ \d(g_{\e^*})\  \prod_{e\not\in\cT}
\d(G_e) \nonumber \\
&=&\int_G \prod_{e^*} dg_{e^*}\
\le(\prod_{\stackindex{e^*\in\cT^*}{\ e^*\neq\e^*}} \d(g_{e^*})
\ri)\ \d(g_{\e^*})\  \prod_{e\not\in\cT} \d(\tG_e).
\label{eqn:effectchange}
\end{eqnarray}
where the elements $\tG_e$ are the elements $G_e$ where the role
of $g_{\e^*}$ and $g_{\te^*}$ are exchanged. Let us prove this
claim by a detailed inspection of the effect of the change of
variable (\ref{eqn:changevariable}) on the RHS of
(\ref{eqn:effectchange}). We examine separately each term and make
a detailed discussion for each type of elements $G_e$.
\begin{itemize}
\item The measure is unaffected by this
change of variable since the Haar measure is invariant by
left/right translation and inverse.
\item The term
\begin{equation}
\prod_{\stackindex{e^*\in\cT^*}{e^*\neq\e^*}} \d(g_{e^*}),
\end{equation}
is unaffected. The reason is that these $e^*$ have either 1) not
any vertex under $v^*$ and $g_{e^*}$ is unaffected, or 2) both
vertices under $v^*$ and $g_{e^*}$ is conjugated into
$k^{-1}g_{e^*}k$ which does not change $\d(g_{e^*})$. The only
edge of $\cT^*$ that has exactly one vertex under $v^*$ is $\e^*$
and it is excluded from this product.
\item $\d(g_{\e^*})$ is unaffected since $g_{\e^*}$ is not
changed.
\item if $e$ is such that $\e^*\not\subset e$ and
$\te^*\not\subset e$, then $G_e$ becomes
\begin{equation}
G_e \to \tG_e=\prodor{e^*\subset e}
\le(k^{-\s(e^*)}g_{e^*}k^{\tau(e^*)}\ri)^{\e(e,e^*)}.
\end{equation}
One can now check that the factors $\s$ and $\tau$ inside the
ordered product exactly compensate, since we are taking the
product along a dual face $f^*$. The only possible effect is to
conjugate $G_e$, which does not affect $\d(G_e)$.
\item if $\e^*\subset e$ and $\te^*\not\subset e$ then $G_e$ becomes
\begin{equation}
G_e \to \tG_e = g_{\e^*} \prodor{\stackindex{e^*\subset e}{e^*\neq
\e^*}} \le(k^{-\s(e^*)}g_{e^*}k^{\tau(e^*)}\ri)^{\e(e,e^*)}
\end{equation}
We choose the starting vertex at $v^*$ and the orientation of the
$e$ that is compatible with $\e^*$. Once again one can use the
fact that the $\s$ and $\tau$ are compensate each other inside the
product. We just to worry about what happens at the beginning and
end of the product. We now have the fact that one of the vertex of
$\e^*$ is $v^*$ and the other is not $\preceq v^*$ by
construction. We thus have
\begin{equation}
\tG_{\e}= g_{\e^*} k^{-1} \le(\prodor{\stackindex{e^*\subset
e}{e^*\neq \e^*}} g_{e^*}^{\e(e,e^*)}\ri) = g_{\te^*}
\prodor{\stackindex{e^*\subset e}{e^*\neq \e^*}}
g_{e^*}^{\e(e,e^*)}
\end{equation}
Then $\tG_e$ is truly $G_e$ with the role of $g_{\e^*}$ and
$g_{\te^*}$ exchanged.
\item if $\te^*\subset e$ and $\e^*\not\subset e$, $G_e$ becomes
\begin{equation}
G_e \to \tG_e = g_{\e^*} g_{\te^*}^{-1} g_{\e^*}
\prodor{\stackindex{e^*\subset e}{e^*\neq\te^*}}
\le(k^{-\s(e^*)}g_{e^*}k^{\tau(e^*)}\ri)^{\e(e,e^*)}
\end{equation}
We use a similar argument than the previous case. Due to the fact
that one of the vertex of $\te^*$ is $v^*$ and the other $R^*$
(hence not under $v^*$) we get
\begin{equation}
\tG_e=g_{\e^*} g_{\te^*}^{-1} g_{\e^*} k^{-1}
\prodor{\stackindex{e^*\subset e}{e^*\neq\te^*}}
g_{e^*}^{\e(e,e^*)}
\end{equation}
which is $G_e$ with the role of $g_{\e^*}$ and $g_{\te^*}$
exchanged.
\item if $\te^*\subset e$ and $\e^*\subset e$, we orient $f^*\sim e$
starting from $v^*$ in a compatible way with $\e^*$, $G_e$ becomes
\begin{equation}
G_e \to \tG_e = g_{\e^*}g_{\e^*}^{-1} g_{\te^*} g_{\e^*}^{-1}
\le(\prodor{\stackindex{e^*\subset e}{e^*\neq\e^*,
\te^*}}k^{-\s(e^*)}g_{e^*}k^{\tau(e^*)}\ri)^{\e(e,e^*)} =
g_{\te^*} g_{\e^*}^{-1} \le(\prodor{\stackindex{e^*\subset
e}{e^*\neq\e^*,\te^*}} g_{e^*}^{\e(e,e^*)}\ri)
\end{equation}
\end{itemize}}
 \underline{Renaming $g_{\e}$ and $g_{\te^*}$ :} The final result
of the investigation is that the $\tG_e$ are the $G_e$ with the
roles of $g_{\e^*}$ and $g_{\te^*}$ exchanged which proves the
claim made in (\ref{eqn:effectchange}). Now in the RHS of this
expression, we can rename $g_{\e^*}$ and $g_{\te^*}$ and get
(\ref{eqn:moveTstar}) which is the move we wanted to perform.

 \vspace{1ex}

\begin{prop}\label{prop:invtree}
The definition of $PR[\cM,\D,\cT,\cT^*]$ does not
depend on the choice of $\cT$.
\end{prop}
\proof We will prove that $PR$ is invariant under an elementary
move of the tree $\cT$. For that we choose a vertex $v$ in $\cT$,
an edge $\e\in\cT$, an edge $\te\not\in\cT$ and we try to prove
the following equality
\begin{equation}
\le(\prod_{e^*} \int_G dg_{e^*} \prod_{e^*\in\cT^*} \d(g_{e^*})
\ri)\ \prod_{e\not\in\cT,e\neq\te} \d(G_e)\
\d(G_{\te})=\le(\prod_{e^*} \int_G dg_{e^*} \prod_{e^*\in\cT^*}
\d(g_{e^*}) \ri)\ \prod_{e\not\in\cT,e\neq\te} \d(G_e)\ \d(G_\e)
\end{equation}
\begin{figure}[t]\psfrag{R}{$R$}\psfrag{ep}{$\e$}\psfrag{te}{$\te$}\psfrag{w}{}\psfrag{v}{$v$}

\includegraphics[width=4cm,height=6cm]{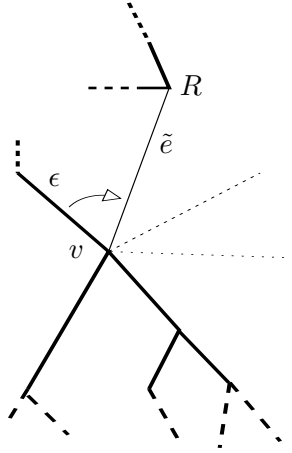}
\caption{Elementary move of $\cT$ at a vertex $v$. The tree is in
thick lines. }\label{fig:movetreeT}
\end{figure}
The root of the tree is chosen to be the other vertex of $e$, and
denoted $R$, see figure \ref{fig:movetreeT}. Here again, we will
need an ingredient which is the fact that if the integrand
contains a $\d(g)$, then the group element $g$ can be inserted
anywhere, since it is imposed to be identity. The property of the
integrand we will need is the following one
\begin{lemma}[Bianchi identity]\label{Bianchid}
For every set of dual faces $f^*$ forming a closed surface $\cS^*$
having the topology of a 2-sphere, it exists some group elements
$k_{f^*}$ (functions of the $g_{e^*}$), some $\s_{f^*}=\pm 1$ and
an order of the faces $f^*$ such that
\begin{equation}
\prodor{f^*\in\cS^*} k_{f^*} G_{f^*}^{\s_{f^*}} k_{f^*}^{-1}=1
\end{equation}

Two particular cases are the following
\begin{itemize}
\item If $\D$ is a triangulation of a real manifold (not a
pseudo-manifold), the neighborhood of every vertex $v$ is a ball,
and the set of $f^*$ dual to the edges $e\supset v$ form a
2-sphere. We then have
\begin{equation}
\prodor{e\supset v} k_{e} G_{e}^{\s_{e}} k_{e}^{-1}=1
\end{equation}
\item If $T$ is a tree of $\D$ (for a real manifold), its tubular
neighborhood has the
topology of a 3-ball and its boundary has the topology of a
2-sphere. This surface can be built has the union of the $f^*$ for
$f^*\sim e$, with $e$ touching the vertices of $T$ without
belonging to $T$.
\end{itemize}
\end{lemma}
\proof Let us consider the dual faces $f^*$ forming a closed
surface $\cS^*$ with the topology of a 2-sphere. It is easy to see
that, if we adapt the starting vertices and orientations , it
exists a way to take the product of the $U_e$ such that the
constitutive $g_{e^*}$ cancels each other and the net result is
just the identity. The elements $k$ and the signs $\s$ express the
difference between starting vertices and orientation we need to
perform this operation, and those chosen for definition of the
model. Let us remark that to write the ordered product, we have a
freedom in the choice the starting dual face $f^*$\ \
$\blacksquare$

\vspace{1ex}

Let us now apply this lemma to the tree $T_v$ which is the subtree
of $\cT$ made by all edges and vertices under $v$ ($v$ is thus the
root of this $T_v$). Choosing the dual face to $\te$ as start, the
Bianchi identity in the neighborhood of $T_v$ can be written as
\begin{equation}
k_{\te} G_{\te} k_{\te}^{-1} = \le(\prod_{e\in S_1} k_e
G^{\s(e)}_e k_e^{-1}\ri) k_\e G_\e k_\e^{-1} \le(\prod_{e\in S_2}
k_e G^{\s(e)}_e k_e^{-1}\ri)
\end{equation}
where the products on the RHS are on edges touching vertices under
$v$, without belonging to $T_1$. These edges are split into two
sets $S_1$ and $S_2$ that we do not need to describe. Due to this
equality, one can write
\begin{equation}
\d(G_{\te})=\le[\le(\prod_{e\in S_1} k_e G^{\s(e)}_e k_e^{-1}\ri)
k_\e G^{\s(\e)}_\e k_\e^{-1} \le(\prod_{e\in S_2} k_e G^{\s(e)}_e
k_e^{-1}\ri)\ri]
\end{equation}
Now all the $G_e$ involved in the RHS are out of $\cT$ and are not
$\te$, thus if we multiply both LHS and RHS by
$\prod_{\stackindex{e\not\in\cT}{e\neq\te}} \d(G_e)$ we get
\begin{equation}
\prod_{e\not\in\cT,e\neq\te} \d(G_e)\ \
\d(G_{\te})=\prod_{e\not\in\cT,e\neq\te} \d(G_e)\ \ \d(G_{\e})
\end{equation}
which achieves the elementary move.\ \ $\blacksquare$

\subsection{Pachner moves}

\begin{prop}\label{prop:invtriang}
The definition of $Z_{PR}[\cM,\D]$ does not depend on $\D$.
\end{prop}
\proof We are now in position to prove that our definition does
not depend on the choice of triangulation. This is done by proving
its invariance under the Pachner moves. We want to consider a
triangulation $\tD$ obtained from $\D$ by Pachner move, and prove
$$
PR(\D)=PR(\tD)
$$
But we need a choice of trees to explicitly write such invariant,
and the previous proposition says the choices are all equivalent.
So what we need to prove is
\begin{equation}
PR(\D,\cT,\cT^*)=PR(\tD,\ctT,\ctT^*)\label{Tree}
\end{equation}
for any choice of trees. The strategy is the following, we start
from a triangulation $\D$ and choose arbitrary trees $\cT$ and
$\cT*$.Performing a move on $\D$ leads to a new triangulation $\tD$, but
we have to specify a choice of trees for this triangulation. We
explicitly described such trees $\ctT$ and $\ctT^*$ as extension
of $\cT$ and $\cT^*$.
It should be noted that we
truly prove the equality (\ref{Tree}) without any additional multiplicative
(infinite) factor that appears under the move and has to be
divided out.

We write all the associated $\d(U_e)$, without taking care
of the fact that some of them will not actually be in the
integrand (for $e\in\cT$) and without taking care of the fact that
certain group elements living on the dual edges will be imposed to
identity by a delta function (those for dual edges in $\cT^*$).
Performing a move on $\D$ leads to a new triangulation $\tD$, but
we have to specify a choice of trees for this triangulation. We
explicitly described such trees $\ctT$ and $\ctT^*$ as extension
of $\cT$ and $\cT^*$. We then prove that
$PR(\D,\cT,\cT^*)=PR(\tD,\ctT,\ctT^*)$. It should be noted that we
truly prove this equality without any additional multiplicative
(infinite) factor that appears under the move and has to be
divided out.

A Pachner move add or remove vertices, edges, faces and
tetrahedra. To prove that $PR(\D,\cT,\cT^*)=PR(\tD,\ctT,\ctT^*)$,
we need to show that
\begin{eqnarray}
&&\le( \int_G \prod_{e^*\in\D^*} dg_{e^*} \prod_{e^*\in\cT^*}
\d(g_{e^*}) \ri)\ \prod_{\stackindex{e\in\D}{e\not\in\cT}} \d(G_e)
\nonumber
\\ &=& \le(\int_G \prod_{e^*\in\tD^*}  dg_{e^*}
\prod_{e^*\in\ctT^*} \d(g_{e^*}) \ri)\
\prod_{\stackindex{e\in\tD}{e\not\in\ctT}} \d(G_e)
\end{eqnarray}
But in this equality, most of the integrals and $\d$-functions are
identical in the RHS and LHS, since the Pachner move affects the
triangulation locally. We just have to consider the objects that
are actually affected by the move, and we just need to prove the
equality of the distributions
\begin{eqnarray}\label{eqn:Pachnerequaldistrib}
&&\le(\int_G \prod_{\stackindex{e^*\in\D^*}{e^*\not\in\tD^*}}
dg_{e^*} \prod_{\stackindex{e^*\in\cT^*}{e^*\not\in\ctT^*}}
\d(g_{e^*}) \ri)\
\prod_{\stackindex{e\in\D,e\not\in\cT}{e\ \ modified}} \d(G_e)
\nonumber \\
&=&\le(\int_G \prod_{\stackindex{e^*\in\tD^*}{e^*\not\in\D^*}}
dg_{e^*} \prod_{\stackindex{e^*\in\ctT^*}{e^*\not\in\cT^*}}
\d(g_{e^*}) \ri)\ \prod_{\stackindex{e\in\tD,e\not\in\ctT}{e\ \
modified}} \d(G_e)
\end{eqnarray}
where ($e\ \ modified$) means that the edge is either removed from
$\D$, or added in $\tD$, or such that $G_e$ is modified by the
move due to faces added or removed. These notations will become
more clear on the specific examples of moves.

\vspace{1ex}

\begin{figure}[t]
\includegraphics[height=6cm]{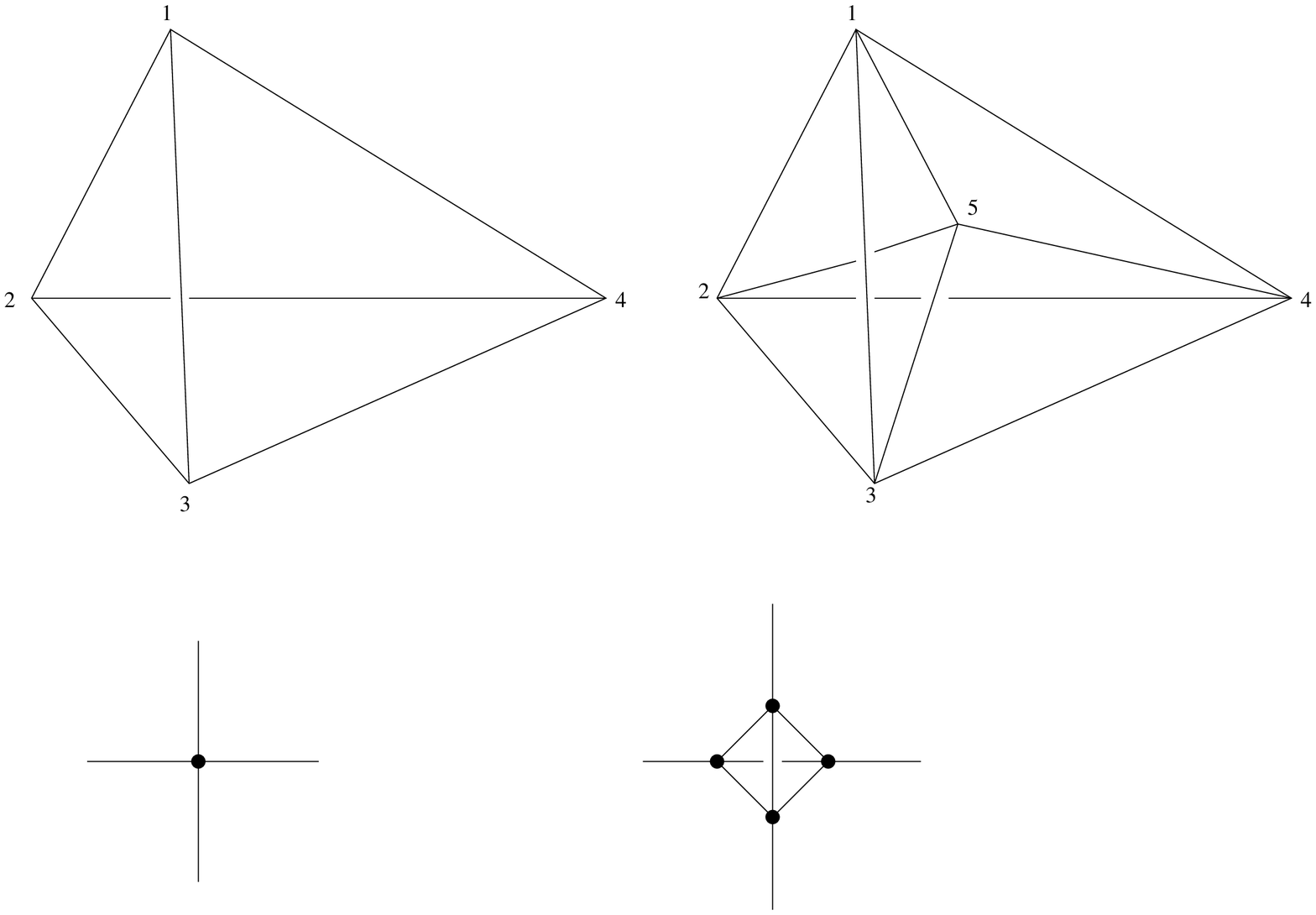}\caption{Pachner moves
$(1\to4)$ and its dual
}\label{fig:Pachner14}
\end{figure}

\underline{Move (1,4) :} Let us first look at the $(1,4)$ move
(see figure \ref{fig:Pachner14}). This move
\begin{itemize}
\item Remove one tetrahedra [(1234)]
\item Add 1 vertex [(5)]
\item Add 4 edges [(15),(25),(35) and (45)]
\item Add 6 faces [(125),(135),(145),(235),(245) and (345)]
\item Add 4 tetrahedra [(1235),(1245),(1345),(2345)].
\end{itemize}
There is one more vertex, this means that we have to add one edge
to $\cT$ to get a $\ctT$, and three more tetrahedra, so we have to
add three edges to $\cT^*$ to get a $\ctT^*$. Let us add the edge
$(15)$ to extend $\cT$ into $\ctT$ and the faces (dual edges)
$(235)$, $(245)$, $(345)$ to extend $\cT^*$ into $\ctT^*$. Let us
explicitly write the LHS and RHS of the equality
(\ref{eqn:Pachnerequaldistrib}) we want to prove. All the edges
represented on the figure are affected. We add some superscript to
the $\d$-functions to make explicit reference to the edge they
come from. We get
\begin{eqnarray}
LHS&=&
\d^{(12)}(g_{124}g_{123}\tG_{12})\d^{(13)}(g_{123}^{-1}g_{134}\tG_{13})
\d^{(14)}(g_{134}^{-1}g_{124}^{-1}\tG_{14})
\nonumber \\ %
&&\d^{(23)}(g_{234}g_{123}\tG_{23})\d^{(24)}(g_{124}g_{234}^{-1}\tG_{24})
\d^{(34)}(g_{234}g_{134}\tG_{34})
\end{eqnarray}
where the $\tG_{ij}$ denote the product of other group elements
for faces containing the edge $(ij)$. They are six more faces
after the move, so 6 more group integrals. They are four new edges
but one of the corresponding $\d$-function is removed due to the
extension of the tree $\cT$ into $\ctT$, so we have 3 more $\d$
functions. We also add three dual edges to get $\ctT^*$ so we have
3 other $\d$-functions. We get for the RHS
\begin{eqnarray}
RHS&=&\int dg_{125}dg_{135}dg_{145}dg_{235}dg_{245}dg_{345}\
\d(g_{235}) \d(g_{245}) \d(g_{345})\nonumber \\
&&\d^{(12)}(g_{124}g_{125}g_{123}\tG_{12})\d^{(13)}(g_{123}^{-1}g_{135}g_{134}\tG_{13})
\d^{(14)}(g_{134}^{-1}g_{145}g_{124}^{-1}\tG_{14}) \nonumber \\
&&\d^{(23)}(g_{234}g_{235}g_{123}\tG_{23})
\d^{(24)}(g_{124}g_{245}g_{234}^{-1}\tG_{24})
\d^{(34)}(g_{234}g_{345}g_{134}\tG_{34})\nonumber \\
&&\d^{(25)}(g_{125}g_{235}^{-1}g_{245}^{-1})
\d^{(35)}(g_{135}g_{345}^{-1}g_{235})
\d^{(45)}(g_{145}g_{245}g_{345})
\end{eqnarray}
Now one can explicitly solve the first line of $\d$-functions,
then the last line and check we get LHS.

\vspace{1ex}

\begin{figure}[t]
\includegraphics[height=6cm]{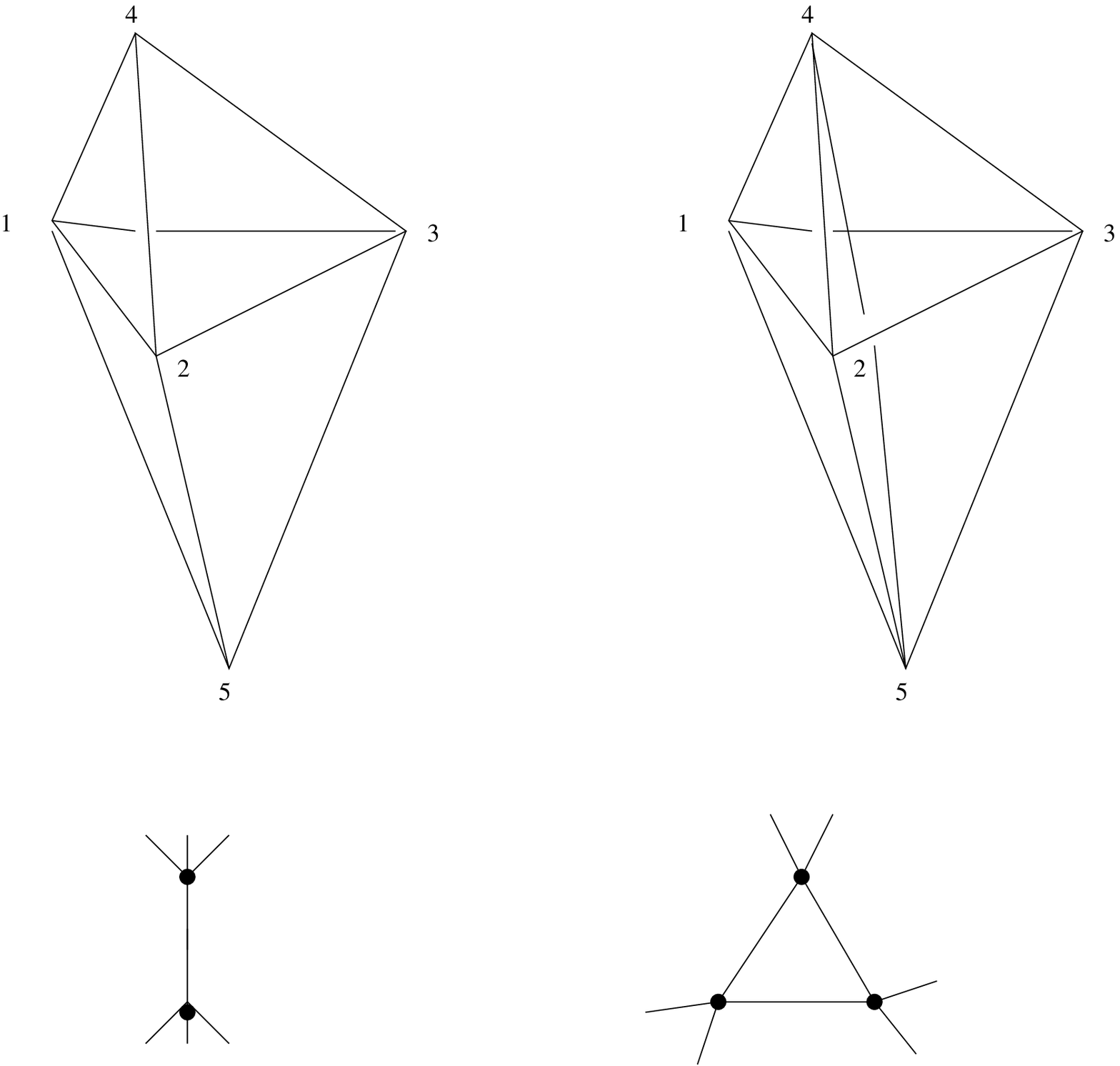}
\caption{Pachner moves $(2\to3)$ and its dual
}\label{fig:Pachner23}
\end{figure}

\noindent\underline{Move (2,3) :} Let us now look at the $(2,3)$
move (see figure \ref{fig:Pachner23}). This move
\begin{itemize}
\item Removes one face (123)
\item Remove two tetrahedra [(1234) and (1235)]
\item Add three tetrahedra [(1245),(1345) and (2345)]
\item Add three faces [(145),(245) and (345)].
\end{itemize}

We have one more tetrahedron so we need to add one edge to $\cT^*$
to get $\ctT^*$. We choose to add (145). The LHS of
(\ref{eqn:Pachnerequaldistrib}) reads
\begin{eqnarray}
LHS &=&\int dg_{123}\ \d^{(12)}(g_{125}g_{123}g_{124}\tG_{12})
\d^{(23)}(g_{235}g_{123}g_{234}\tG_{23})
\d^{(13)}(g_{134}g_{123}^{-1}g_{135}\tG_{13}) \nonumber \\
&&\d^{(14)}(g_{124}^{-1}g_{134}^{-1}\tG_{14})
\d^{(24)}(g_{234}^{-1}g_{124}\tG_{24})
\d^{(34)}(g_{134}g_{234}\tG_{34}) \nonumber \\
&&\d^{(15)}(g_{135}^{-1}g_{125}^{-1}\tG_{15})
\d^{(25)}(g_{125}g_{235}^{-1}\tG_{25})
\d^{(35)}(g_{235}g_{135}\tG_{35})
\end{eqnarray}
The RHS is
\begin{eqnarray}
RHS &=& \int dg_{145}dg_{245}dg_{345} \d(g_{145}) \\ &&
\d^{(12)}(g_{125}g_{124}\tG_{12})
\d^{(23)}(g_{235}g_{234}\tG_{23})
\d^{(13)}(g_{134}g_{135}\tG_{13}) \nonumber \\
&&\d^{(14)}(g_{124}^{-1}g_{145}g_{134}^{-1}\tG_{14})
\d^{(24)}(g_{234}^{-1}g_{245}g_{124}\tG_{15})
\d^{(34)}(g_{134}g_{345}g_{234}\tG_{34}) \nonumber \\
&&\d^{(15)}(g_{135}^{-1}g_{145}^{-1}g_{125}^{-1}\tG_{15})
\d^{(25)}(g_{125}g_{245}^{-1}g_{235}^{-1})
\d^{(35)}(g_{235}g_{345}^{-1}g_{135}\tG_{35})
\d^{(45)}(g_{145}g_{345}g_{245})
\end{eqnarray}
We first solve $\d(g_{145})$. We can then solve $\d^{(45)}$ and
get $g_{245}=g_{345}^{-1}$, we will call $g_{123}$ this common
group element for purpose of comparison. We get
\begin{eqnarray}
RHS &=& \int dg_{123} \d^{(12)}(g_{125}g_{124}\tG_{12})
\d^{(23)}(g_{235}g_{234}\tG_{23})
\d^{(13)}(g_{134}g_{135}\tG_{13}) \nonumber \\
&&\d^{(14)}(g_{124}^{-1}g_{134}^{-1}\tG_{14})
\d^{(24)}(g_{234}^{-1}g_{123}g_{124}\tG_{15})
\d^{(34)}(g_{134}g_{123}^{-1}g_{234}\tG_{34}) \nonumber \\
&&\d^{(15)}(g_{135}^{-1}g_{125}^{-1}\tG_{15})
\d^{(25)}(g_{125}g_{123}^{-1}g_{235}^{-1})
\d^{(35)}(g_{235}g_{123}g_{135}\tG_{35})
\end{eqnarray}
One can now use the following translation
\begin{eqnarray}
g_{125} &\to&  g_{125}g_{123} \\ g_{234} &\to& g_{123}g_{234} \\
g_{135} &\to&  g_{123}^{-1}g_{135}
\end{eqnarray}
and get the LHS.

\ \ $\blacksquare$

\end{document}